\documentclass[fleqn, 12pt]{wlscirep}
\usepackage{subfig}
\usepackage[utf8]{inputenc}
\usepackage[T1]{fontenc}
\usepackage{amssymb}
\usepackage[font=footnotesize,labelfont=bf,
   justification=justified,
   format=plain]{caption} 

\title{Addressing Data Quality Challenges in Observational Ambulatory Studies: Analysis, Methodologies and Practical Solutions for Wrist-worn Wearable Monitoring}

\author[1,*, 
]{Jonas Van Der Donckt}
\author[2, 3]{Nicolas Vandenbussche}
\author[1]{Jeroen Van Der Donckt}
\author[1]{Stephanie Chen}
\author[1]{Marija Stojchevska}
\author[1]{Mathias De Brouwer}
\author[1]{Bram Steenwinckel}
\author[2, 3]{Koen Paemeleire}
\author[1]{Femke Ongenae}
\author[1]{Sofie Van Hoecke}

\affil[1]{IDLab-imec, Faculty of Engineering and Architecture, Ghent University}
\affil[2]{Department of Neurology, Ghent University Hospital}
\affil[3]{Department of Basic and Applied Medical Sciences, Faculty of Medicine and Health Sciences, Ghent University}


\begin{abstract}
Chronic disease management and follow-up are vital for realizing sustained patient well-being and optimal health outcomes. Recent advancements in wearable sensing technologies, particularly wrist-worn devices, offer promising solutions for longitudinal patient follow-up by shifting from subjective, intermittent self-reporting to objective, continuous monitoring. However, collecting and analyzing wearable data presents unique challenges, such as data entry errors, non-wear periods, missing wearable data, and wearable artifacts. We therefore present an in-depth exploration of data analysis challenges tied to wrist-worn wearables and ambulatory label acquisition, using two real-world datasets (i.e., mBrain21 and ETRI lifelog2020). We introduce novel practical countermeasures, including participant compliance visualizations, interaction-triggered questionnaires to assess personal bias, and an optimized wearable non-wear detection pipeline. Further, we propose a visual analytics approach to validate processing pipelines using scalable tools such as tsflex and Plotly-Resampler. Lastly, we investigate the impact of missing wearable data on “window-of-interest” analysis methodologies. Prioritizing transparency and reproducibility, we offer open access to our detailed code examples, facilitating adaptation in future wearable research. In conclusion, our contributions provide actionable approaches for wearable data collection and analysis in chronic disease management.
\end{abstract}
\begin{document}

\flushbottom
\maketitle
%
%
\thispagestyle{empty}


\section*{Introduction}

In recent years, wearable sensing has seen a vast increase in both research and commercialization, driven by the reduced networking snort d hardware costs as well as the non-intrusive nature of these devices \cite{heikenfeld_wearable_2018}. Wearable sensing offers promising solutions for patient monitoring by continuously acquiring objective physiological data in an unobtrusive manner and at scale. As such, wearable sensing could potentially ease the strain on the healthcare system, particularly in the management of chronic diseases \cite{baig_systematic_2017}. For instance, diabetes patients could benefit from real-time tracking of blood sugar levels through wearable sensing and timely intervention \cite{klonoff_continuous_2005}. Similarly, patients with cardiovascular conditions might use wearable sensors to monitor vital signs, providing early detection of anomalies and enabling prompt medical attention \cite{bayoumy_smart_2021}. Wearable sensing empowers individuals to actively manage their health and reduces the burden on healthcare by preventing complications and hospitalizations associated with chronic diseases \cite{taylor_does_2021}.

To effectively implement remote monitoring, one needs to integrate data entries from patients and/or healthcare providers with data from wearable sensors obtained in ambulatory settings \cite{rodgers_recent_2015}. This integration necessitates evaluating the wearable’s ability to detect specific events, such as fall detection for the elderly \cite{chen_wearable_2006}, or identifying and validating biomarkers in real-life settings \cite{kim_wearable_2019}. Furthermore, remote monitoring has proven valuable in tracking and analyzing chronic events in certain populations, such as headache attacks of migraine patients \cite{de_brouwer_mbrain_2022, siirtola_using_2018, stubberud_forecasting_2023}, seizures in epilepsy patients \cite{bottcher_detecting_2021}, or depressive episodes \cite{heikenfeld_wearable_2018}. These studies underline the potential of wearable sensing for chronic disease follow-up and management \cite{baig_systematic_2017}.

Given the potential of remote monitoring, there are an increasing number of studies that collect wearable data along with acute event data in ambulatory settings. However, substantial analytical challenges emerge when conducting such monitoring studies in real-world scenarios \cite{schmidt_wearable-based_2019}. These data quality challenges manifest across different phases, spanning participant acquisition, data collection, and retrospective analysis, and encompass issues related to participants, monitoring devices and applications, and technologies \cite{canali_challenges_2022, cho_factors_2021}. Many studies in ambulatory wearable monitoring currently either overlook or sidestep these challenges \cite{siirtola_using_2018, stubberud_forecasting_2023, liao_future_2019}. In this paper, we aim to address this gap by providing actionable countermeasures to the below identified data quality challenges.

Specifically, we categorize the data quality challenges into two groups: those related to (i) participants and monitoring applications, and (ii) wearables, thereby excluding other domain challenges such as technology or scalability. The (i) participant and monitoring application category includes problems with data entry and quality, lack of participant compliance and motivation, assumptions made without clear evidence, and personal biases. The (ii) wearable category focuses on user wearable non-wear, artifacts from the wearables themselves, and missing wearable data.

To address (i) participant and application-related challenges, we introduce a novel participant compliance visualization technique to monitor participant motivation in a near-real-time manner. In addition, we propose interaction-triggered questionnaires, designed to aid in reducing data entry errors. Regarding the second category, (ii) wearable analysis challenges, we introduce an efficient non-wear detection pipeline, allowing processing wearable data at larger scales. Moreover, we propose a generic visual analytics approach to validate such signal processing pipelines. Lastly, we outline a bootstrapping technique devised to assess the impact of metric computation on partially missing data segments.
To elucidate these challenges, we draw upon our first-hand experience during the mBrain study \cite{de_brouwer_mbrain_2022}. By utilizing an excerpt of this \href{https://www.kaggle.com/datasets/jonvdrdo/mbrain21/data}{mBrain21 dataset} along with the ETRI lifelog 2020 dataset~\cite{chung_realworld_2022}, we substantiate our proposed countermeasures with \href{https://github.com/predict-idlab/data-quality-challenges-wearables}{reproducible implementations}. As such, this work aims to aid future monitoring studies in bridging the gap between recognizing the occurrence of the identified challenges and the practical applicability of countermeasures.

\section*{Related Work}
Over the past decade, research interest in ambulatory wearable-based monitoring studies has significantly increased, leading to several works partially indicating challenges and limitations inherent to such studies. In this section, we outline works that consider these challenges.
In 2018, Schmidt et al.~\cite{schmidt_labelling_2018} formulated guidelines and practical implementation details, derived from their field study, for enhancing the accuracy of manual data entries for ecological momentary assessments (EMA) in the context of ambulatory wearable monitoring studies. Their recommendations emphasize the importance of brevity in the duration of each EMA, along with ensuring that the EMA only targets the core goal of the study. Additionally, they advocate for daily screenings of the wearable’s signal modalities, facilitating researchers to routinely assess data quality. This is crucial to perform timely re-instructions to participants when a decline in data quality becomes apparent. Furthermore, in order to minimize the intrusiveness of manual data entry, they suggest that participants should also have the option to configure the EMA application to match their circadian rhythm. To sustain participant engagement, they suggest leveraging strategies such as incremental reward systems. Unfortunately, practical examples of a participant intervention or re-instruction are not present. Moreover, this work does not focus on providing methodologies to leverage collected ambulatory wearable data, along with EMA events, in downstream analysis.

In Schmidt et al. (2019)~\cite{schmidt_wearable-based_2019}, the authors expand on these EMA-related guidelines, integrating wearable data processing. However, this work does not specifically focus on the wearable data challenges associated with field studies.

Balbim et al. (2021)~\cite{balbim_using_2021} discussed ambulatory data quality challenges associated with FitBit Physical Activity (PA) trackers. Their work mainly concerns study preparation, intervention delivery, and study closeout, whilst leaving out data analysis.

The systematic review of Cho et al. (2021)~\cite{cho_factors_2021} identifies three overarching factors that influence wearable data quality: device- and technical-related factors, user-related factors, and data governance-related factors. Within their taxonomy, device- and technical-related factors include hardware issues such as sensor degradation or malfunction, software issues related to the quality of proprietary wearable algorithms, and networking issues. User-related factors encompass instances of wearable non-wear and user errors stemming from wearable misplacement or poor skin contact. Lastly, data governance-related factors largely arise from the absence of standardization in data formats, inconsistency in algorithms across different devices, and variations in sensor placements. The review of Cho et al. primarily concentrates on elucidating these factors without emphasizing methodologies to address them. Aligning with their taxonomy, we aim to address user-related and device-related factors, thereby omitting technical and data governance challenges.

In recent work, Böttcher et al.~\cite{bottcher_data_2022} assessed the data quality of the Empatica E4 wearable, particularly in the context of epilepsy monitoring, using multiple datasets from hospitalized and ambulatory care settings. The data quality was evaluated by computing signal quality scores for several physiological signal modalities of the Empatica, namely skin conductance (EDA), blood volume pulse (BVP), and skin temperature (TMP). Furthermore, an on-body score was utilized to analyze non-wear periods, along with a data completeness score, representing the ratio between the actual recorded volume and expected data volume. Their findings suggested superior data quality for physiological modalities during nighttime (8PM-8AM) compared to daytime. Furthermore, nighttime data showed greater data completeness, likely due to fewer disconnections during data streaming. Notably, streaming revealed a higher data loss compared to on-device logging. Across all datasets, high on-body scores (>80\%) were observed. The study by Böttcher et al. was centered on assessing and comparing wearable data quality in both hospitalized and ambulatory settings, and while it highlighted the data quality superiority of nighttime data, the work was not scoped to offer actionable methodologies for conducting analysis nor improving data quality during study collection. Moreover, while the authors made their analysis results publicly available (\href{https://github.com/WEAR-ISG/WEAR-DataQuality/tree/main}{GitHub}), no documentation was provided, except for the main analysis file. Unfortunately, due to data-sharing agreements for the patient cohorts, they could not share the source data publicly.

In conclusion, while considerable research has indicated data quality challenges prevalent in wearable monitoring studies, there remains a notable gap in providing actionable countermeasures, tangible examples, and streamlined code for addressing post-hoc analysis issues. Furthermore, few of these studies grant access to their data or code, complicating the assessment of their methodologies' broader applicability.

\section*{Methodology}
This section outlines our approach to tackling data quality challenges. First, we introduce two distinct datasets employed to demonstrate these challenges and highlight their characteristics. Next, we define the scope of our work, distilling the specific data quality challenges we aim to address. Finally, we describe the programming environment and tools chosen to tackle these challenges.

\subsection*{Datasets}

We materialize data quality challenges by using examples from the mBrain21 and ETRI lifeLog 2020 datasets, whose characteristics are outlined in Table~\ref{tab:1}.

\begin{table}[h]
\footnotesize
\centering
\begin{tabular}{@{}rcc@{}}
\toprule
\multicolumn{1}{l}{}        & \textbf{mBrain21*} & \textbf{ETRI lifelog 2020} \\ \midrule
Subjects                    & 4*                 & 22                         \\
Country                     & Belgium            & South Korea                \\
Age range (median, 95\% CI) & \textit{/}         & 28 {[}21, 33{]}            \\
Sex (\% female)             & \textit{/}         & 41\%                       \\
Study duration              & 90 days            & 28 days                    \\
Wearable type               & Empatica E4        & Empatica E4                \\
Wearable placement          & Wrist              & Wrist                      \\
Recording mode              & streaming          & device                     \\
Labeling                    & self-report        & self-report                \\ \midrule
\multicolumn{3}{l}{*this dataset only provides an excerpt of 4 participants}  \\ \bottomrule
\end{tabular}
\caption{Comparative overview of the characteristics of the two datasets selected for this work.}
\label{tab:1}
\end{table}

The mBrain study involved monitoring patients diagnosed with chronic headache disorders over a 90-day period. This monitoring was conducted using smartphone sensors (i.e., movement, application usage), and the Empatica E4 wrist-worn wearable, along with a dedicated application to record their headache events, medication intake, and respond to daily questionnaires~\cite{van_der_donckt_self-reporting_2022}. Moreover, participants were instructed to wear the Empatica device for at least 8 hours per day. The Empatica E4 was connected to a dedicated application that streamed the wearable data to internal servers after buffering for two minutes. This near real-time data stream was then utilized to construct automatic timelines of activity and stress predictions, offering an interactive timeline interface to the participants, as demonstrated in Figure 1. The primary objective of the mBrain study was to meticulously analyze ambulatory wearable data in relation to headache intervals. For instance, wearable movement data can be utilized to investigate if there is a change in movement behavior during cluster headaches, as demonstrated in Vandenbussche et al. (2023)~\cite{vandenbussche_patients_2024}.

\begin{figure}[!hbt]
\centering
    \subfloat[\centering]{{\includegraphics[width=0.45\linewidth]{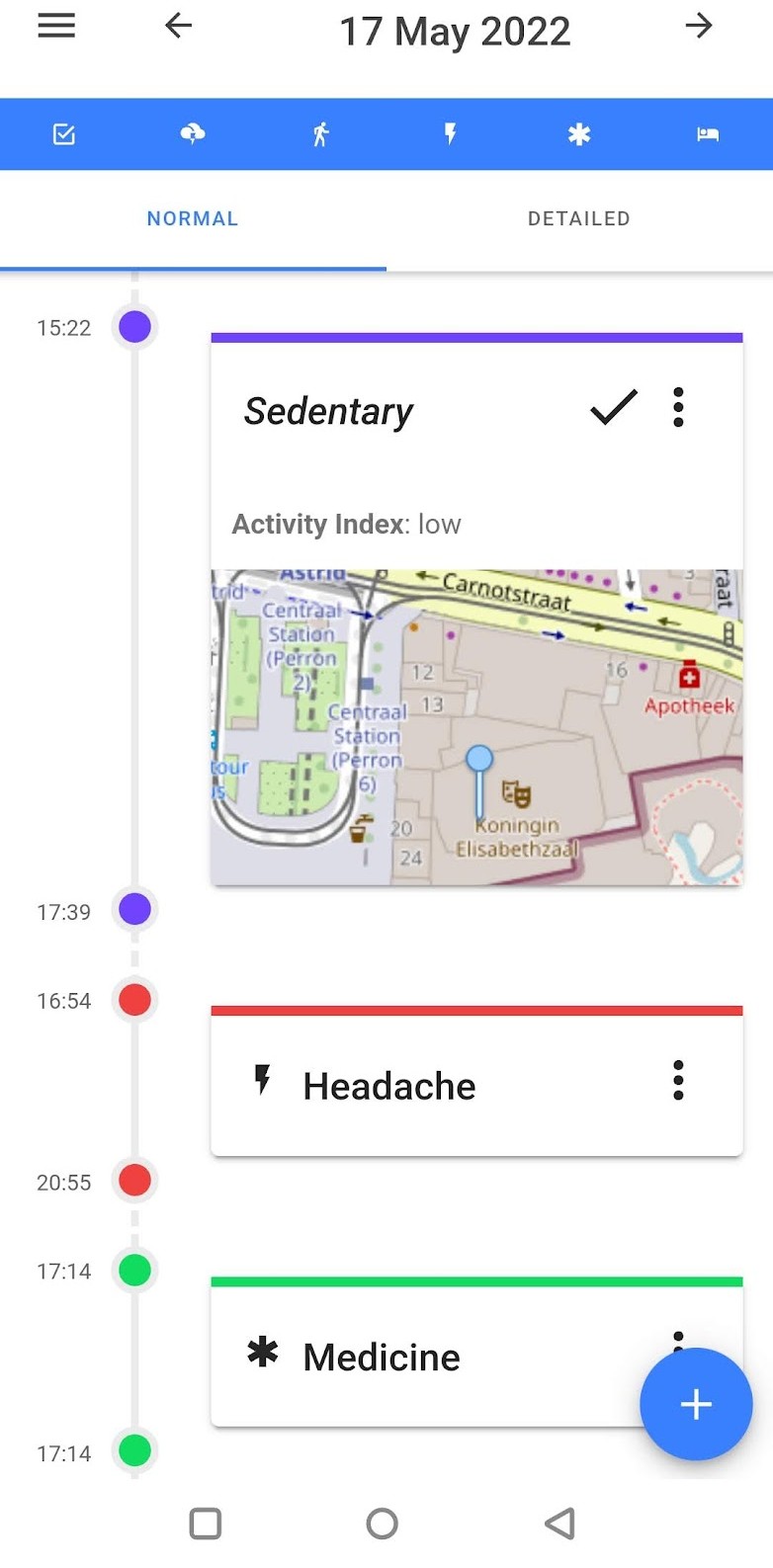}}}
    \qquad
    \subfloat[\centering]{{\includegraphics[width=0.45\linewidth]{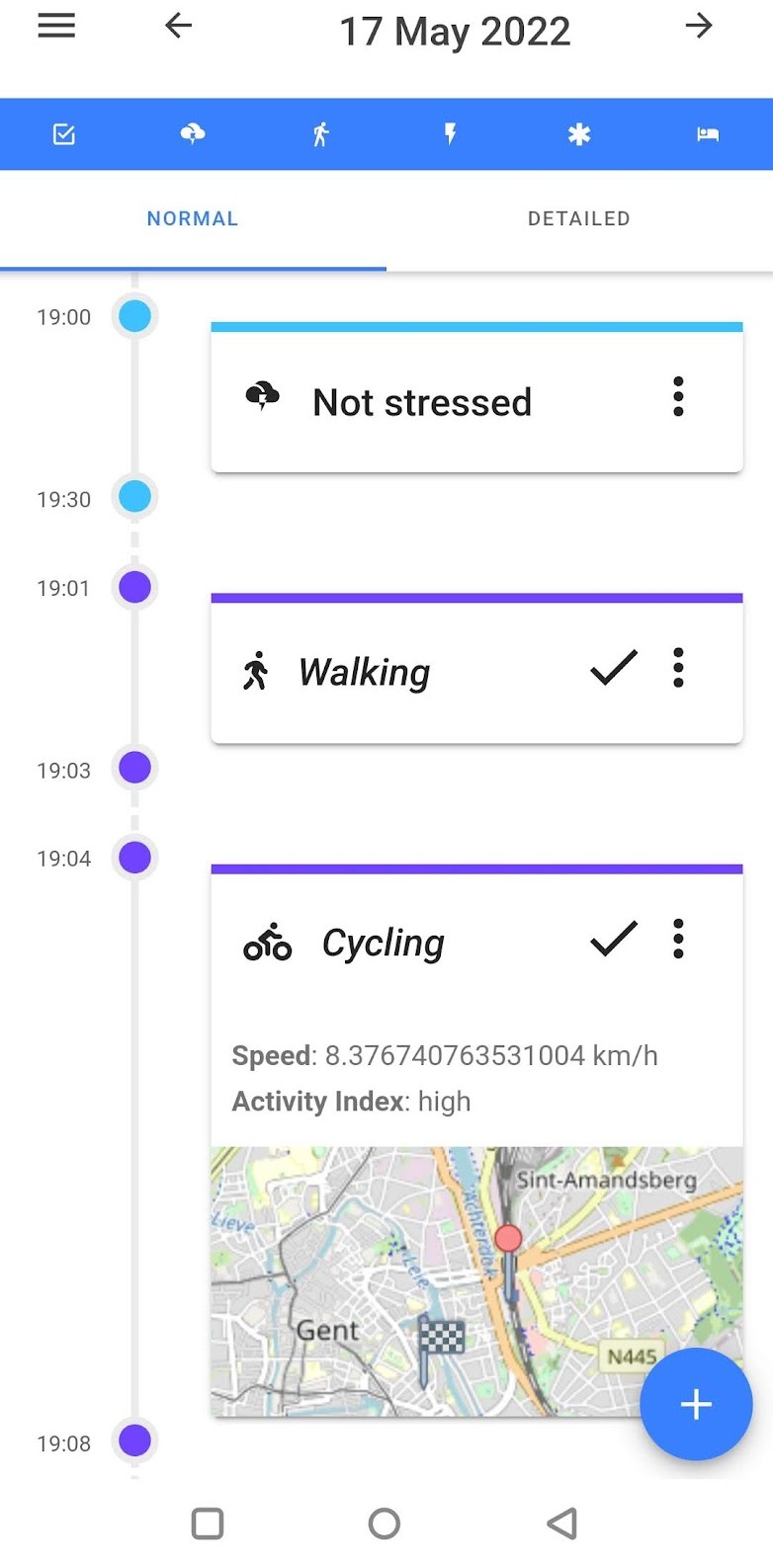}}}
     \caption{mBrain application timeline of a dummy participant, showcasing contextual data, including user-defined semantic locations (e.g. “Work”).}
\label{fig:timeline}
\end{figure}

The ETRI lifelog 2020 study monitored 22 participants for 28 days with the aim of acquiring data-driven descriptions of human life for various perspectives~\cite{chung_realworld_2022}. Specifically, the ETRI dataset is composed of a smartphone, a wearable (Empatica E4), and a sleep-quality monitoring sensor (Withings sleep tracking mat). Participants utilized a dedicated application which facilitated self-reporting of their activity, social state (alone, with someone, with a group), along with their semantic location (e.g., home, work), and experienced emotional state (valence, arousal). These self-reported labels were also presented to the user via a timeline. Participants were also tasked with offloading the on-device logged Empatica data to a computer which then uploads it to Empatica’s cloud.

Both studies under consideration utilized the Empatica E4, a medically graded wristband that captures physiological and movement data. The E4 contains a three-axis accelerometer which samples at 32 Hz with a range of +/-2 g. The 64 Hz blood volume pulse (BVP) signal is constructed from a proprietary on-device algorithm that leverages the green and red exposure photoplethysmography (PPG) signals~\cite{empatica_srl_e4_nodate}. This derived BVP signal serves as input for proprietary algorithms which compute the inter-beat-interval (IBI) timings and the mean heart rate (HR). The IBI timings are computed directly on the Empatica device, but only when the BVP data are considered of adequate quality. In contrast, the mean HR is calculated using an algorithm on the Empatica Cloud platform and the results are consistently made available. The skin surface temperature (TEMP) is acquired at 4 Hz via a thermopile sensor. Lastly, the skin conductance or electrodermal activity (EDA) is acquired at 4 Hz via two AgCl electrodes.

We deliberately selected these two datasets given our direct experience with the mBrain21 dataset and the well-documented nature and availability of the ETRI lifelog 2020 dataset. Since both datasets are recorded by different research institutes, and capture different demographic populations, we believe that they should demonstrate a certain genericity of our presented methodologies. Due to our lack of direct involvement during the ETRI lifelog study phase, we primarily use examples from the mBrain study to illustrate the challenges of the participant data entry category.

\subsection*{Selecting Data Quality Challenges}
The focus of this work lies on challenges related to data completeness and correctness of ambulatory monitoring studies, with a specific emphasis on those employing wrist-worn devices along with an application for ambulatory label acquisition. The specificity of our focus largely stems from first-hand experience with the mBrain project, which falls under this study category. Moreover, this study type is frequently employed by smaller-scale studies to assess the wearables’ potential of detecting these ambulatory labeled events-of-interest, such as affect, headaches, and stress \cite{siirtola_using_2018, stubberud_forecasting_2023, schmidt_wearable-based_2019}.

We categorize the data quality challenges into two domains: (1) participant data entry challenges, and (2) wearable analysis challenges. The first category, participant data entry challenges, encompasses the participant and application-oriented challenges that impact data quality. These include participant compliance and motivation (challenge 1; C1), implicitness assumptions (C2), data entry errors (C3), and personal bias (C4). Conversely, the wearable data challenges concentrate on wearable-related analysis challenges, including wearable non-wear (C5), wearable artifacts (C6), and the analysis of “windows-of-interest” featuring missing and anomalous data (C7). For each of these seven identified challenges, we provide detailed insights into their causes, impacts, and potential countermeasures. Wherever possible, we illustrate these countermeasures with concrete visualizations and implementation examples on either the data analysis side (retrospective) or the application side (prospective or reactive).

\subsection*{Programming Environments}
Longitudinal wearable monitoring studies produce large datasets. As indicated by the Kaggle 2022 data-science survey, notebook-based environments, particularly those using IPython, are the go-to tools for data scientists~\cite{kaggle_kaggle_2022}. Interactive notebook-based formats drive data exploration, which is crucial in every step of the data science process~\cite{perkel_why_2018}. Consequently, this study employs IPython environments to illustrate methods for tackling data-centric challenges. The Python packages utilized in this work are listed and managed using the Poetry Python package manager~\cite{pypoetry_poetry_nodate}.

Notably, both Weed et al. (2022)~\cite{weed_impact_2022} and Böttcher et al. (2022)~\cite{bottcher_data_2022} utilized MATLAB to perform their wearable analyses. However, we believe that Python’s scalability, open-source nature, larger community, and easier integration with other technologies, along with cost-effectiveness and flexibility, is a more suitable choice.

\section*{Participant data Entry Challenges}
In this section, human and application-related data entry challenges of wearable monitoring studies are covered.

\subsection*{Challenge 1: Participant Compliance}
Participant motivation to wear a wearable device, interact with questionnaires, and submit experienced events of interest is crucial to obtain a qualitative dataset. However, it has been demonstrated in numerous cases, including our own mBrain study~\cite{de_brouwer_mbrain_2022}, that participant motivation, and thus also study interaction, tends to decrease over time~\cite{schmidt_labelling_2018, heger_using_2023}. This compliance decrease can be attributed to multiple components, such as an inconvenient wearable connection process, wearable design aesthetics, too frequent EMA collection, inconvenient EMA moments, or poor study user experience (including adverse reactions to the wearable device). In order to minimize the user burden, each labor-intensive component during the study period, including wearable wear requirements and EMAs, should target the core goal of the research, while including as little overhead as possible~\cite{schmidt_labelling_2018, muaremi_towards_2013}. Moreover, frequent or long-term manual data entry can lead to response fatigue, which may diminish data quality and accuracy~\cite{porter_multiple_2004}.

\begin{figure}[!hbt]
\centering
    \includegraphics[width=\linewidth]{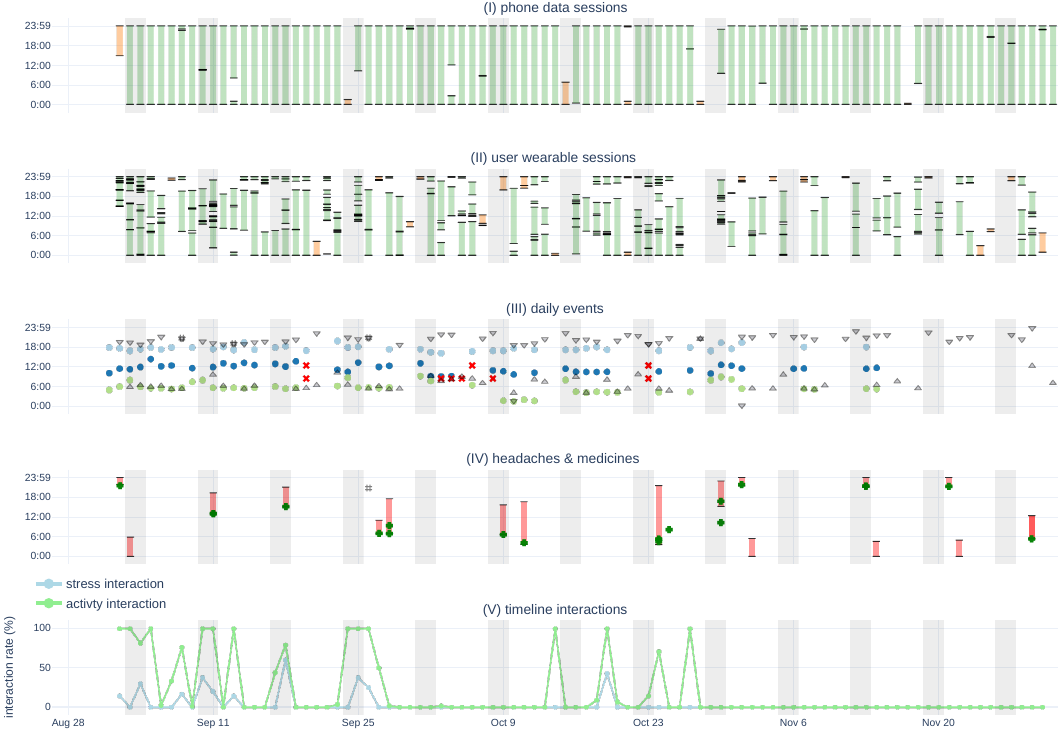}
     \caption{mBrain study interaction visualization of a single participant for a period of 90 days. \textit{Note}: The figure consists of several subplots with a shared x-axis, each providing different layers of information about the participant's activity and interactions. Subplots (i) and (ii) showcase phone and wearable data sessions over time. Each bar on the x-axis represents a unique day. For the first four plots, the y-axis indicates the time of day. This format reveals patterns of data fragmentation and daily volumes across extended periods. Gray-shaded areas, consistent across all subplots, signify weekends. The mBrain study requires a minimum of eight hours of wearable data daily. This compliance is color-coded in the first two subplots: green represents days with more than 8 hours, while orange indicates less than 8 hours. The daily events subplot (iii) provides an overview of food intakes (dot-shaped markers; green: breakfast, blue: lunch, light blue: dinner, red-cross: skipped meal) and questionnaire interaction ($\nabla$: evening questionnaire, $\triangle$: morning questionnaire, \#: stress event questionnaires). Subplot (iv) provides a visual record of the participant's headaches (depicted by red vertical bars) and medication intakes (indicated by green crosses). The final subplot (v), which denotes an interaction rate (\%) as y-axis, illustrates the frequency of participant interactions with stress (light blue) and activity (light green) timeline events derived from the wearable data stream.}
\label{fig:study_interaction}
\end{figure}

Furthermore, it is crucial to continuously ensure participant commitment throughout the study, as keeping subjects motivated will ensure high-quality data and labels, regarding both frequency and completeness~\cite{schmidt_labelling_2018}. This can be achieved through an incremental reward system or gamification~\cite{hutchison_out_2010, ottenstein_compliance_2022, van_berkel_gamification_2017}. Additionally, periodic contact with participants throughout the study has been demonstrated as an effective strategy to uphold motivation~\cite{gloster_daily_2017}. As a reactive counteraction, it is also worthwhile to query participants during study takeout on their experiences, feelings, and gains from participating, as this could give insights into issues arising from such long ongoing participant interaction.

\subsubsection*{Countermeasures}
From a data science perspective, several solutions can be devised to address the aforementioned challenges. During the mBrain study, we created a novel methodology to evaluate the continuity of participant motivation by generating interaction rate reports derived from the incoming data streams. Furthermore, integrating real-time notifications, such as through webhooks linked to a messaging platform, can be utilized to notify study coordinators when participants do not meet the studies' interaction threshold.

\begin{figure}[!hbt]
\centering
    \includegraphics[width=\linewidth]{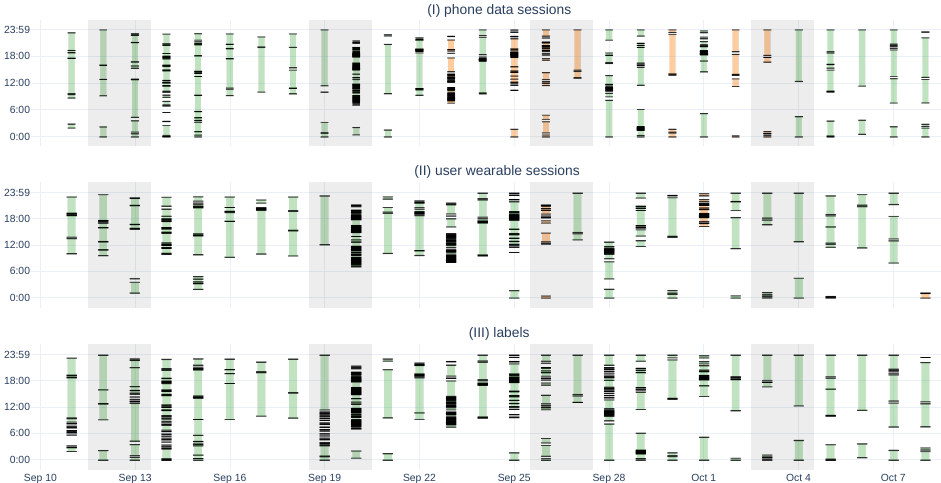}
     \caption{ETRI lifelog 2020 study interaction visualization of a single participant for a period of 28 days. \textit{Note}: Similar to the mBrain interaction plot (Figure~\ref{fig:study_interaction}), the ETRI interaction visualization utilizes stacked bars to depict phone (i) and wearable (ii) data sessions. Accommodating the fact that participants manually labeled intervals, the label subplot (iii) also uses a bar interval representation, indicating periods for which social and affective labels are present. Remark how the phone and wearable session are a subset of the label session data. In alignment with Figure~\ref{fig:study_interaction}, session bars are color-coded in orange when fewer than 8 hours of data are available for the corresponding day.}
\label{fig:etri_study_interaction}
\end{figure}

\begin{figure}[!hbt]
\centering
    \includegraphics[width=0.6\linewidth]{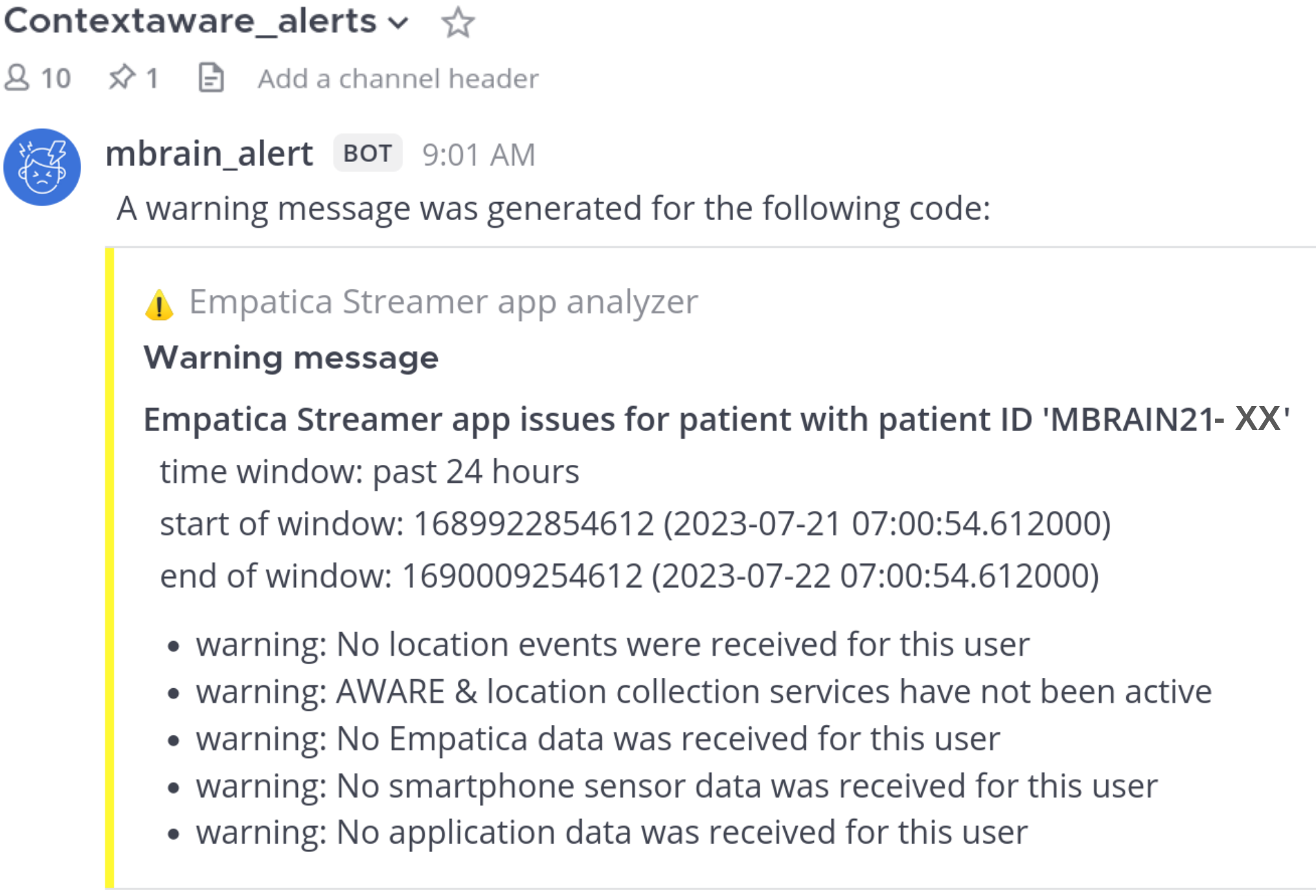}
     \caption{Example of an mBrain alert message, shown to the study coordinators when no wearable data is received from a participant.}
\label{fig:mbrain_webhook_notifs}
\end{figure}

Figure~\ref{fig:study_interaction} illustrates this novel participant compliance methodology applied to the mBrain study. Specifically, we utilized scheduled jobs to generate these daily participant compliance reports. Participant compliance is assessed by displaying the intervals for which there is wearable and phone data, the evolution of questionnaires answered, etc. This visualization allows study coordinators to have an overview of the participant’s study compliance, facilitating participant re-instruction in a timely manner when needed. To showcase the generalizability of our proposed method, we also applied this interaction rate visualization to the ETRI lifelog 2020 dataset, as shown in Figure~\ref{fig:etri_study_interaction}.

In the mBrain study, the participant compliance reports were also complemented by real-time alert messages, as shown in Figure~\ref{fig:mbrain_webhook_notifs}, which notified study coordinators when an active mBrain participant had streamed less than 8 hours of data in the last 24 hours. 

Based on both participant compliance assessment methodologies (reports and alert messages) one can decide whether participants need to be contacted (e.g., via a phone call, or personalized messages) in order to gauge the reasons for the reduced study compliance. Within the mBrain study, the clinician-neurologist sent - whenever needed, and gardening not to overload participants as well - personalized messages to the participants via a dedicated interface. This methodology was empirically evaluated during the mBrain study and physicians were able to track the interactions of the participant with a compliance visualization dashboard and contact them if a certain threshold of underperformance was met. By doing so, the study design had a built-in monitoring system with rapid notification possibility to enhance study compliance.

\subsection*{Challenge 2: Implicitness Assumptions}
A common assumption in data collection is that the absence of a recorded event implies it did not occur~\cite{de_brouwer_mbrain_2022, fox-wasylyshyn_handling_2005}. For instance, if no medication event is logged for a day, it might be assumed that the patient did not take any medication, but the participant may have simply overlooked logging the event. Therefore, it is essential to implement checks for such implicit assumptions, often through direct questioning about these (absent) self-reported events. 

\begin{figure}[!hbt]
\centering
    \subfloat[\centering]{{\includegraphics[width=0.4\linewidth]{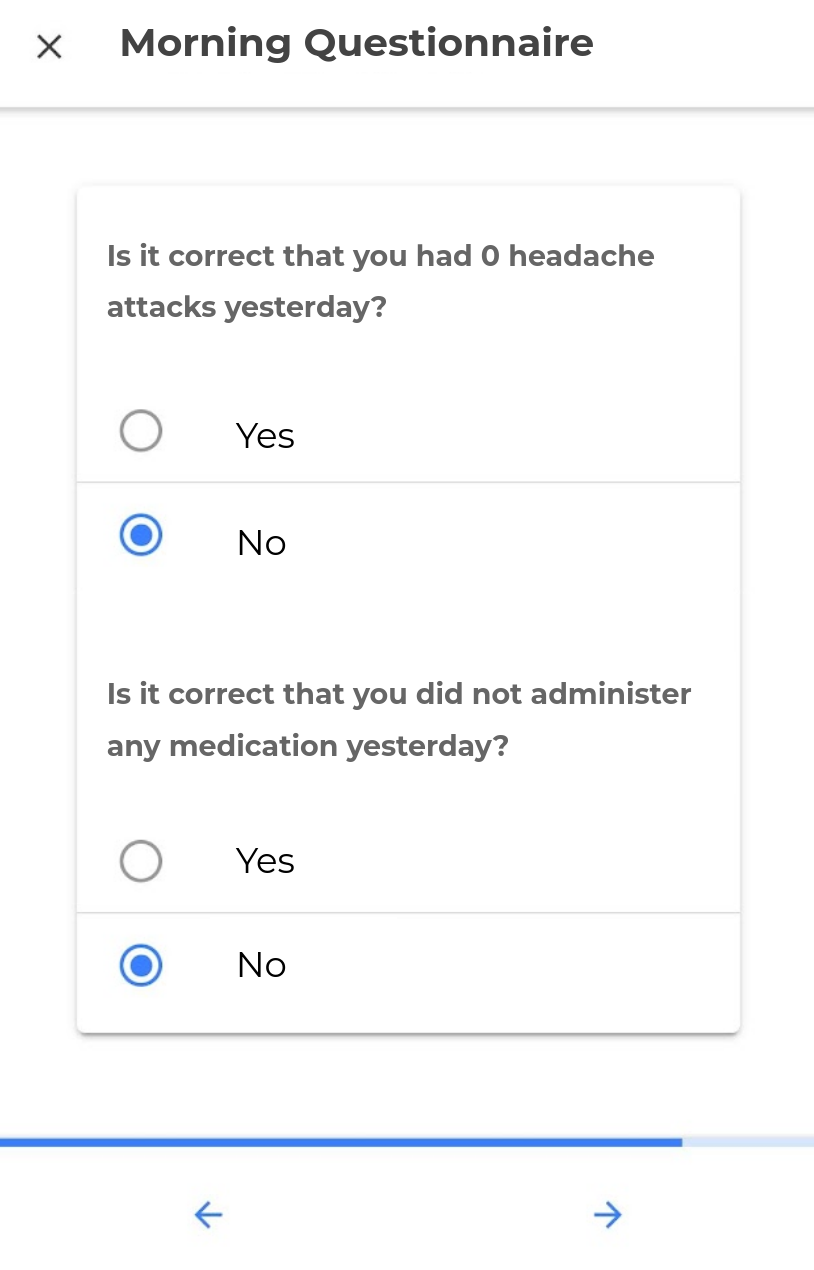}}}
    \qquad
    \subfloat[\centering]{{\includegraphics[width=0.5\linewidth]{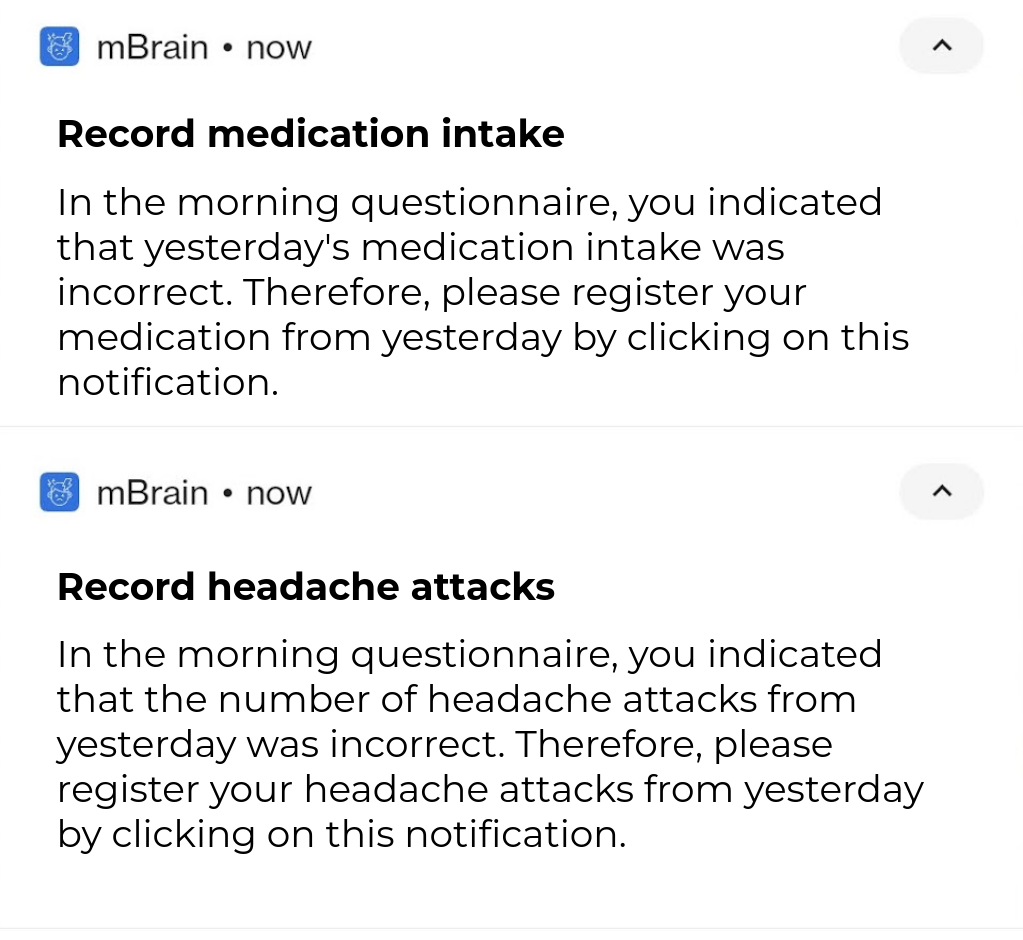}}}
     \caption{(a) Screenshot of questions in the mBrain study’s morning questionnaire evaluating implicitness for headache and medication events. (b) Notifications activated based on responses to the implicitness questions}
\label{fig:implicitness}
\end{figure}

\subsubsection*{Countermeasures}
In the mBrain study, morning questionnaire responses serve to validate or disprove assumptions regarding the occurrence of headaches and medication use of the previous day, as illustrated by Figure~\ref{fig:implicitness} (a). Subplot (b) of Figure~\ref{fig:implicitness} displays notifications sent to the user when their responses contradict these implicit assumptions; i.e., when the user replies “No” to one of the questions in (a). Remark that while these additional queries enhance data accuracy, they also increase the burden on participants. Therefore, these checks should be confined to parameters predominant to the study's analysis. Notably, regularity in questionnaires may improve study compliance by creating a routine for the participant to interact with the study environment, and therefore possibly nudge the participant toward other components of the study~\cite{colls_patient_2021}.

\subsection*{Challenge 3: Data Entry Errors}
Data entry errors often arise from accidental user mistakes during interactions, largely attributed to suboptimal design choices~\cite{baig_mobile_2015}. As such, enhancing user experience through cognitively ergonomic designs can significantly reduce these human errors~\cite{walsh_human_2002}. A proactive strategy is to conduct a pilot phase, involving a small group of participants and analysts. This allows for the identification and correction of issues related to implicitness assumptions and data entry issues prior to full-scale monitoring~\cite{balbim_using_2021}.

Temporal inaccuracies, a specific category of data entry errors, primarily stem from users’ uncertainty in allocating exact timestamps to events~\cite{schmidt_labelling_2018}. These inaccuracies are typically influenced by two biases: recall bias (misdating past events) and predictive bias (misdating future events). Notably, EMAs emerged as a strategy to evaluate immediate experiences in participants’ everyday settings, thereby minimizing recall bias~\cite{csikszentmihalyi_validity_2014}. Moreover, temporal accuracy can be enhanced by integrating additional contextual data, such as location and activities, into an automated timeline, which helps to counteract recall bias~\cite{hoelzemann_matter_2023}.

\subsubsection*{Countermeasures}
\begin{figure}[!hbt]
\centering
    \includegraphics[width=0.45\linewidth]{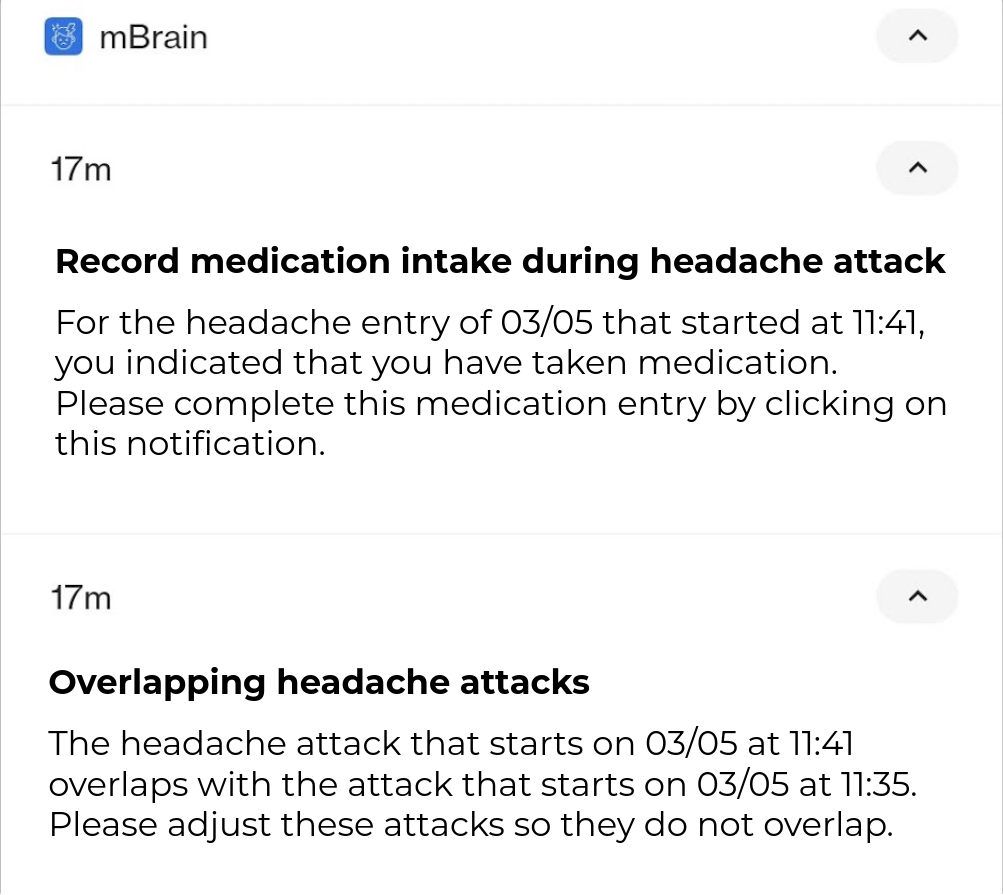}
     \caption{Example mBrain application notifications when conflicting data entries were made by a participant.}
\label{fig:notifs_conflicting_data_entries}
\end{figure}

Within the mBrain study, we conducted two pilot phases to factor out data entry error challenges and assess the robustness of our infrastructure in managing higher user-loads~\cite{bracke_design_2021}. We also propose to utilize an extensive intake procedure. During this intake, participants can review all the components of the application and the wearable device along with the study coordinator. This does not only clarify the process but also benefits the participant’s motivation to perform data entries. During this intake, it is preferred to also provide a manual to the participant, in which all the intricacies of the application and study procedure are described 9. The intake procedure was successfully validated within the mBrain study. Lastly, a reactive measure that we performed, based on the pilot study errors, consists of implementing sanity checks to further reduce data entry errors. This sanity check system prevents patients from logging multiple concurrent events of the same type and generates alerts for entries with improbable dates, such as logging a headache that occurred two weeks in the past or is set for a future date. Figure~\ref{fig:notifs_conflicting_data_entries} showcases notifications in the mBrain study that are triggered by users’ incomplete or ambiguous data entries.

Tackling temporal inaccuracies, we propose to use an automated user timeline, as employed within the mBrain study, sourced from smartphone and wearable data, as shown in Figure~\ref{fig:timeline}. This timeline assists in improving the temporal specificity when pinpointing stress or headache occurrences. Another countermeasure is allowing participants to specify a time range instead of a single, definitive timestamp. This approach recognizes and accommodates the user's temporal uncertainty, such as by letting them denote a span for both the start and end time of a headache event. However, this flexibility might complicate the user experience and eventual analysis, so it should be carefully aligned with the study's objectives to ensure its analytical value.

\subsection*{Challenge 4: Personal Bias}
In remote health monitoring studies, where manual data entries from participants play a central role, personal bias emerges as a substantial challenge. This bias stems from the subjective nature of self-reported data and the various ways in which individual perceptions, beliefs, and motivations influence these reports~\cite{schmidt_labelling_2018}. Addressing personal bias in such studies necessitates a multifaceted approach. Rigorous study design, participant education (to, e.g., homogenize definitions of concepts such as stress), regular reminders, intuitive technology interfaces, and the integration of objective monitoring tools can all play a role in mitigating the effects of this bias. Nonetheless, researchers should always be aware of personal biases when interpreting subjectively labeled data. As such, considering the participant as a latent factor during analysis, by for instance modeling the participant as a random effect with a Linear Mixed Model, is a recommended practice~\cite{sun_utility_2022}. Including the participant as a random effect allows for the modeling of the variability between participants and helps in accounting for the within-subject correlation due to repeated measurements on the same participant. This way, any inherent individual bias or subject-specific characteristic (like baseline levels) that might affect the outcome variable can be taken into consideration during analysis.

Beyond self-reporting, personal bias can also manifest in device wear behavior. For instance, during the mBrain study, we noted that participants wore the wearable devices less frequently during headache episodes (Figure~\ref{fig:wearable_wear_bheavior}). It might also be the case that certain participants opt to not wear the device during more intensive activities, this way possibly skewing the findings. A general countermeasure to this challenge is to instruct participants that the monitoring study aims to observe each aspect of their daily life, and that they therefore must keep wearing the wearable whenever possible. Moreover, when such behavior is observed through compliance reports, participants can also be contacted during the study period (see C1).

\begin{figure}[!hbt]
\centering
    \includegraphics[width=\linewidth]{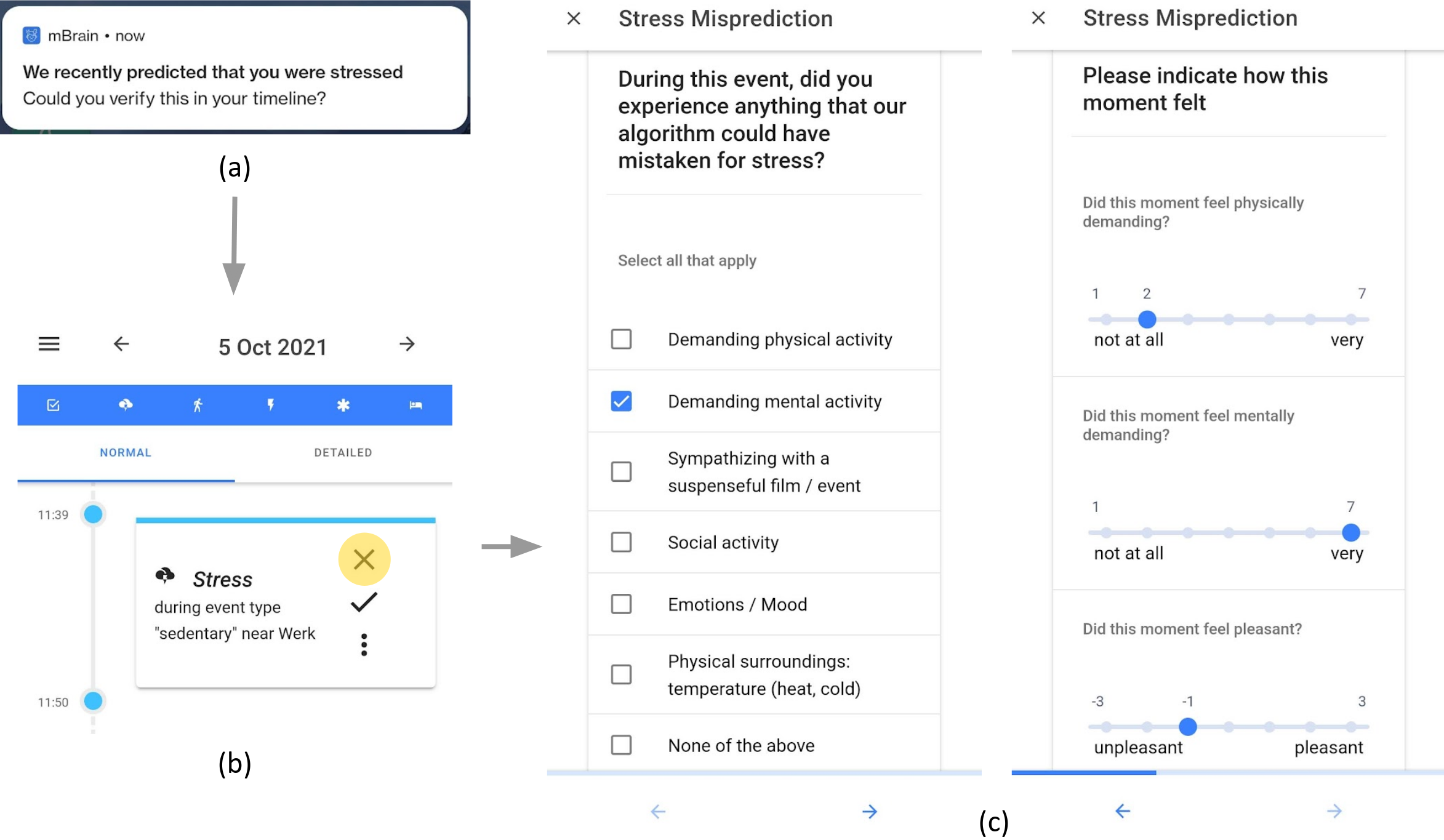}
     \caption{mBrain stress event interaction and its corresponding misprediction questionnaire.\textit{Note}: When a stress-system-activation (e.g., a sudden non-activity induced increase in skin conductance responses) is detected in the streamed wearable data, a notification is sent to the user as shown in (a). This notification aids in reducing the interaction latency of the participant. When clicking on this notification, the participant is guided toward the mBrain timeline in which the recent stress event is shown, as depicted by (b). The yellow circle indicates that the participant re-labeled the stress-event period to be non-stressful. This, in turn, prompts the participant whether they have time to fill in a questionnaire that gauges for more contextual information about this event. This questionnaire is portrayed in (c) and indicates that the user was performing a demanding mental activity which was not perceived as really pleasant, possibly explaining the stress response.}
\label{fig:stress_misprediction}
\end{figure}

Finally, the Hawthorne effect, i.e., the modification of participant behavior in response to being observed, can also affect data representativity~\cite{mombers_identifying_2016}. It has been demonstrated that the Hawthorne effect appears to last for a limited time~\cite{vaisman_out_2020}. Hence, monitoring studies that have an adequate duration (i.e., up to 3-6 months) can mitigate this issue.

\subsubsection*{Countermeasures}
We propose to leverage interaction-triggered questionnaires to gauge for causes and contextual information related to the wrongful or correct prediction of highly personal events such as stress. Figure~\ref{fig:stress_misprediction} (c) depicts such a questionnaire as employed within the mBrain study. This figure demonstrates how - although the participant disproved the stress prediction - the event was perceived as mentally demanding and slightly unpleasant, which might be indicated as stress by other users.

\begin{figure}[!hbt]
\centering
    \includegraphics[width=\linewidth]{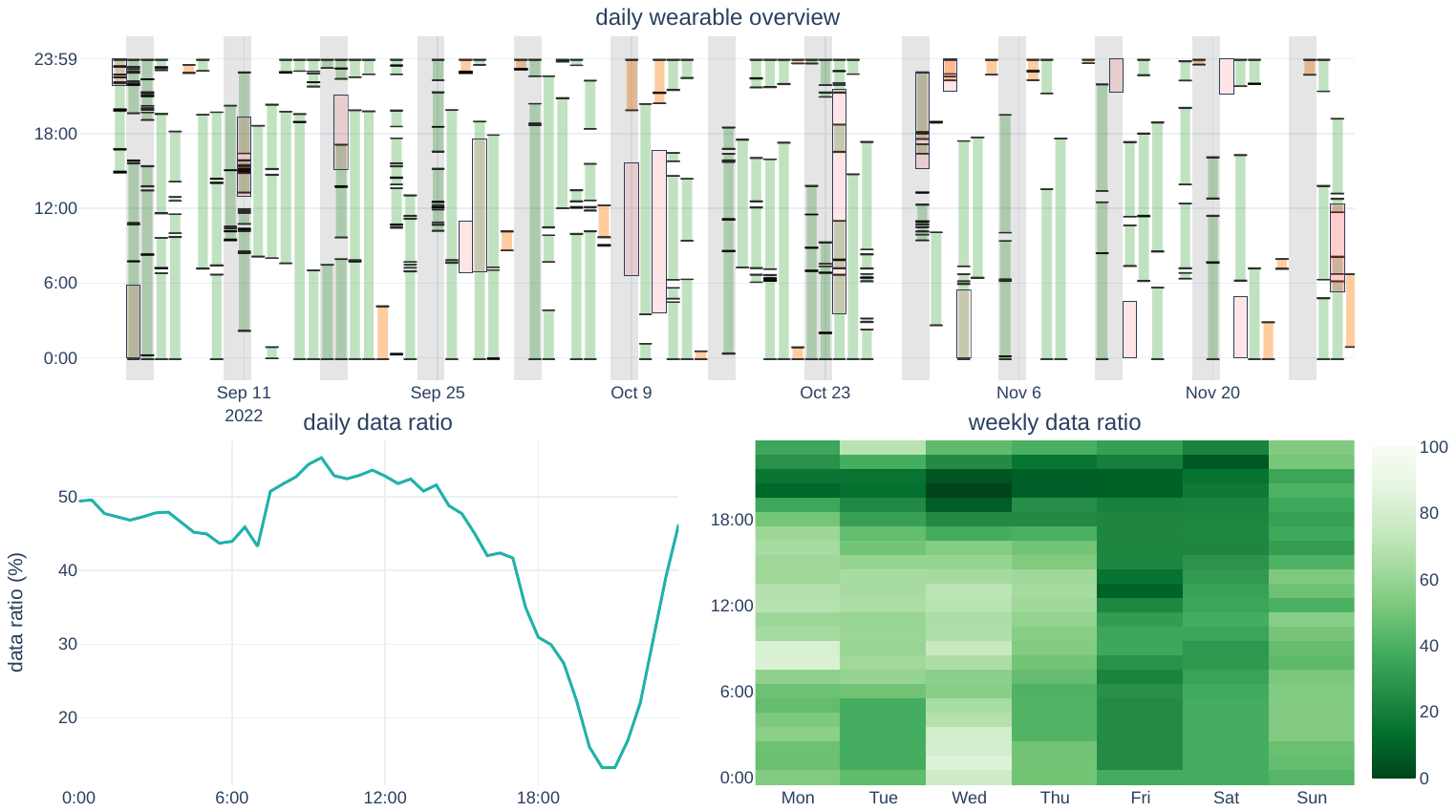}
     \caption{ mBrain study wearable wear behavior overview of a single participant.\textit{Note}:The upper subplot illustrates the available wearable sessions, using similar bar intervals as Figure 2, providing an overview of wearable usage. In this subplot, weekends are marked in gray and headache intervals in red. This participant has an average wearable data ratio of 44\%, whereas the available data ratio during headaches is 39\%. The lower left subplot depicts the average data ratio for the time of day throughout the study period. This subgraph reveals a notable decline in wearable use between 17h30 and 22h30. Conversely, the lower right subplot utilizes a heatmap to display the average data ratio against the time of day, distributed over each day of the week, highlighting discernible patterns in wear frequency. This heatmap elucidates that this specific participant has a tendency for reduced wearable use on Fridays and Saturdays, while Wednesdays exhibit the most wearable use. Remark how the reduced wearable usage during the evening period, shown by the through in the lower left subplot, is also discernable in this heatmap visualization}
\label{fig:wearable_wear_bheavior}
\end{figure}

Moreover, participant interaction visualizations, as illustrated by Figure~\ref{fig:wearable_wear_bheavior} for the mBrain study, allow study coordinators to observe whether participants demonstrate certain trends in non-wear and/or non-activity (e.g., no wearable data during evening periods). The obtained insights can then be utilized to send custom messages to participants regarding their behavior.

\subsection*{Summary}
To conclude this participant data entry section, Table~\ref{tab:summary_data_entry} summarizes the four identified participant and data entry related challenges and their corresponding countermeasures.

\begin{table}[h]
\resizebox{\textwidth}{!}{%
\begin{tabular}{@{}rl@{}}
\toprule
\multicolumn{1}{l}{\textbf{Challenge}} & \textbf{Countermeasures / Actions} \\ \midrule
\multicolumn{1}{l}{\textbf{Participant Compliance}} &  \\
High user burden \& response fatigue & \begin{tabular}[c]{@{}l@{}}App interactions: Minimize overhead; each component should target the core goal\\ Wearable: Strive for a convenient wearable experience (e.g., connection process, battery life, …)\end{tabular} \\
Decline in motivation & \textit{\begin{tabular}[c]{@{}l@{}}(reactive) Monitor participant compliance (+ reinstruct when needed)\\ (proactive) Periodic contact with study coordinator\\ (proactive) Incremental reward systems \& gamification\\ (retrospective) Query experiences during takeout to pinpoint motivation hurdles\end{tabular}} \\ \midrule
\multicolumn{1}{l}{\textbf{Implicitness Assumptions}} &  \\
Event absence assumption & (proactive) Utilize daily questionnaires to validate these assumptions (should always relate to core goal) \\ \midrule
\multicolumn{1}{l}{\textbf{Data Entry Errors}} &  \\
Application Entry Errors & \textit{\begin{tabular}[c]{@{}l@{}}(proactive) Pilot phase to factor out errors\\ (reactive) Sanity checks \& notifications\end{tabular}} \\
Temporal inaccuracies & \textit{\begin{tabular}[c]{@{}l@{}}(proactive) Providing contextual data reduces recall bias\\ (proactive) Gauge for temporal certainty\end{tabular}} \\ \midrule
\multicolumn{1}{l}{\textbf{Personal Bias}} &  \\
Labeling bias & \textit{\begin{tabular}[c]{@{}l@{}}(reactive) Gauge for contextual information / reasoning\\ (retrospective) Include participant effect during analysis\end{tabular}} \\
Wear behavior & \textit{\begin{tabular}[c]{@{}l@{}}(proactive) Instruct participants to wear the device all the time\\ (reactive) Monitor wear behavior and interfere when needed\end{tabular}} \\
Hawthorne effect & Monitor for a sufficient duration (e.g., 3-6 months) \\ \bottomrule
\end{tabular}%
}
 \caption{: Summary of the participant data entry challenges and their countermeasures.}
 \label{tab:summary_data_entry}
\end{table}

\section*{Wearable Analysis Challenges}
In this section, we address the challenges related to ambulatory wearable data quality, in the context of performing data analysis. The objective of such data analysis is primarily centered on examining “windows of interest”, which could entail event-related time-spans (for instance, those obtained by an accompanying application, such as headache periods) or specific intervals of the day (like nighttime). We identify three key challenges: 1) data streaming when the wearable is not on the body (C5), 2) artifacts introduced by the wearable device (C6), and 3) strategies for analyzing wearable data that includes missing or spurious data (C7), resulting from scenarios like non-wear, absence of the device on the body, and device-generated artifacts.

\subsection*{Challenge 5: Wearable not on body}
The issue of non-wear, i.e., intervals during which the wearable continues to record data despite not being worn, has long been acknowledged as a crucial challenge in actigraphy research~\cite{berger_methodological_2008, choi_development_2012}. To address this, a variety of approaches have been established, which utilize wearable movement (ACC) signals for detecting non-wear. This detection is often performed as a preprocessing step, filtering the data before further analyses. Ahmadi et al. (2020)|~\cite{ahmadi_non-wear_2020} evaluated five non-wear detection algorithms which used only wrist-worn accelerometer data. They observed that algorithms based on the standard deviation of the wearable acceleration can effectively detect non-wear periods that last at least 30 minutes.

Recently, there has been an increase in the development of non-wear detection algorithms that incorporate physiological parameters, such as skin temperature and skin conductance in addition to wearable movement. Vert et al. (2022)~\cite{vert_detecting_2022} utilized the GENEActiv wrist-worn wearable, which includes a near-body temperature sensor along with a light sensor. By utilizing the rate-of-change of the temperature signal, their algorithm is able to detect non-wear periods for intervals as short as 5 minutes. Remark that this high temporal specificity is unattainable when exclusively using movement signals. Vert et al. also emphasized the importance of detecting such shorter non-wear periods in free-living scenarios, which often include short removals, e.g., when showering or washing hands. Similarly, Pagnamenta et al. (2022)~\cite{pagnamenta_putting_2022} integrated temperature data into their non-wear detection algorithms for the Axivity AX3 wrist-worn wearable. They used a relative temperature threshold of 3°C to identify non-wear periods for 5-minute windows. Their methodology displayed high sensitivity and specificity compared to algorithms relying solely on accelerometer data. Lastly, Böttcher et al. (2022)~\cite{bottcher_data_2022} developed an on-body score for the Empatica E4, which combines the skin conductance, skin temperature, and movement signals. This binary on-body score is computed for 1-minute intervals and serves to assess data quality in retrospective datasets.

\subsubsection*{Countermeasures}
Given that both datasets under consideration in this work utilize the Empatica E4, and previous studies indicated enhanced accuracy when incorporating physiological signal modalities, we make a novel contribution that refines the algorithm proposed by Böttcher and colleagues to be more efficient and sensitive. Table~\ref{tab:tab_non_wear} provides a side-by-side comparison of the parameter values of both algorithm versions.

\begin{table}[htb]
\centering
\footnotesize
\begin{tabular}{@{}rcc@{}}
\toprule
\multicolumn{1}{l}{\textbf{}} & \textbf{Bottcher et al.} & \textbf{Refined (ours)} \\ \midrule
\textbf{Movement SQI} & \begin{tabular}[c]{@{}c@{}}ACC-SD sum (SD window = 10s)\\ \textgreater{}= 0.2g\end{tabular} & \begin{tabular}[c]{@{}c@{}}ACC\_x-SD (SD window = 1s)\\ \textgreater{}= 0.1 g\end{tabular} \\
Skin temperature SQI & 25 °C \textless{}= valid \textless{}= 40 °C & \textgreater{}= 32 °C \\
Skin conductance SQI & \textgreater{}= 0.05 $\mu S$ & \textgreater{}= 0.03 $\mu S$ \\ \midrule
\textbf{SQI processing} & \begin{tabular}[c]{@{}c@{}}1-minute mean per SQI\\ \textgreater{}= 1\% on-body -\textgreater valid\end{tabular} & \begin{tabular}[c]{@{}c@{}}Reindexing\\ (i.e., ensuring shared index)\end{tabular} \\
SQI aggregation & OR-aggregation & \begin{tabular}[c]{@{}c@{}}OR-aggregation\\ smoothing\end{tabular} \\ \midrule
\textbf{Inference} & 38 ms per hour(*) & 6 ms per hour(*) \\
Granularity & 1 minute & 0.25 seconds \\ \midrule
\multicolumn{3}{l}{\begin{tabular}[c]{@{}l@{}}(*) Both inference timings were computed on the same hardware. \\ A reference notebook with both implementations and timing details can be found here.\end{tabular}} \\ \bottomrule
\end{tabular}
   \caption{ Algorithmic and parameter-based comparison of two non-wear algorithms.}
    \label{tab:tab_non_wear}
\end{table}

\begin{figure}[htp]
    \subfloat[Our refined non-wear algorithm]{\includegraphics[clip,width=\columnwidth]{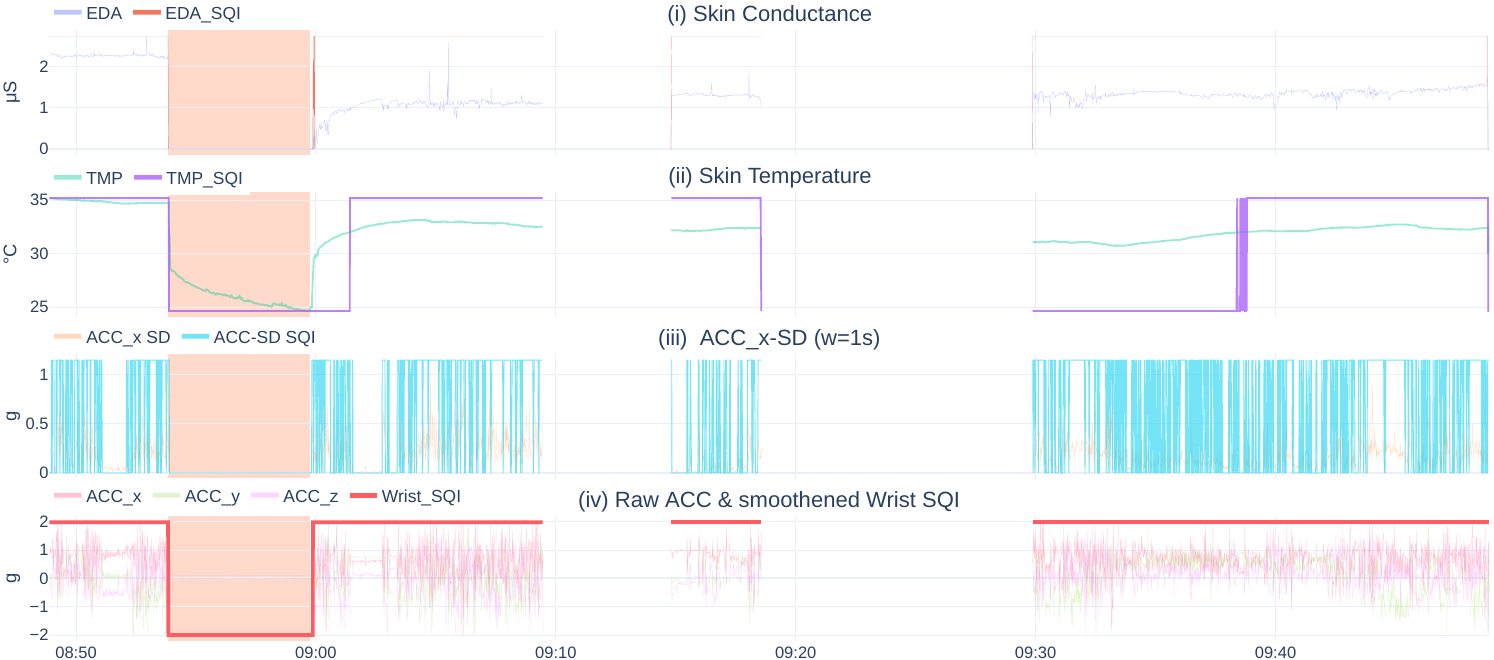}}
    \hspace{0px}
    \subfloat[Non-wear algorithm of Böttcher et al.~\cite{bottcher_data_2022}]{\includegraphics[clip,width=\columnwidth]{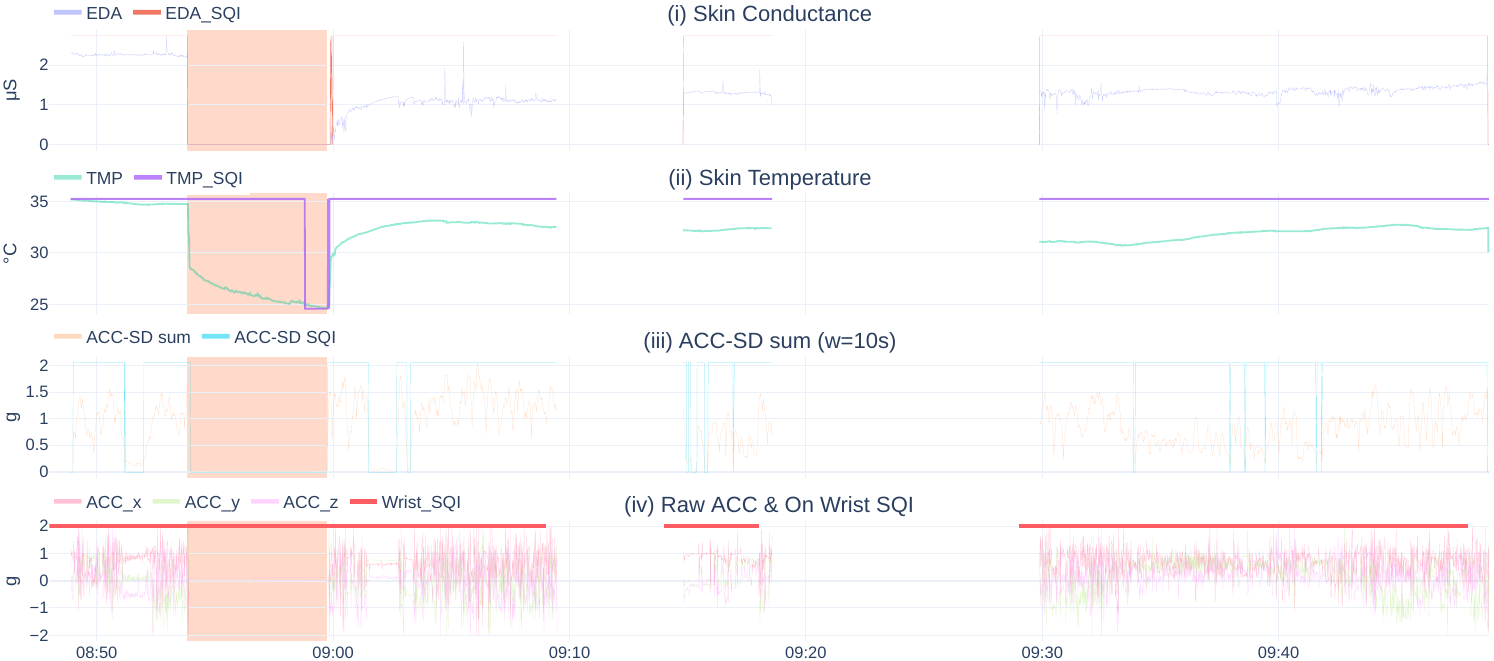}}
    \caption{Visual comparison of Böttcher’s and our refined non-wear detection algorithm on the same excerpt. \textit{Note}: The red-shaded area in each subplot of both (a) and (b) represents a labeled non-wear interval. Subplot (i) and (ii) in figure (a) and (b) depict the signal-specific SQIs for the skin conductance and temperature, while subplot (iii) represents the standard deviation of the ACC and the corresponding ACC-SD SQI. Subplot (iv) shows the three-axis accelerometer data alongside the resulting “Wrist\_SQI”. A low Wrist\_SQI value between 08h55 and 09h00 in figure (a) denotes non-wear. Examining this time-interval in subplots (i) and (ii) of (a), a notable decline in skin conductance and temperature is observed, leading to low SQI values. Minimal movement within this interval also reflects a low SQI value in subplot (iii). Conversely, in figure (b), this non-wear bout remains undetected, primarily due to the valid temperature SQI range (i.e. between 25 °C and 40 °C). This lower bound may be set too low, as only the last part of the skin temperature segment during this non-wear period results in a low SQI.
}
    \label{fig:off_wrist_algorithm}
\end{figure}

Specifically, we simplify the computation of the movement standard deviation signal by only considering the x-axis of the ACC signal, as opposed to calculating the standard deviation for all three axes followed by a summation. This simplification improves the efficiency, as ACC-based operations proved to be the bottleneck of Böttcher’s algorithm, given its high sample rate (32 Hz). Remark that by utilizing a smaller window size for the sliding window SD computation, i.e., 1 second instead of 10 seconds, the efficiency of this computation is further enhanced. Empirical validation indicated a high correlation between the simplified and the original ACC-SD signal. Next on, in alignment with the work of Böttcher, the ACC-SD signal is transformed into a binary Signal Quality Index (SQI) by using a threshold value. Both versions employ thresholding for skin conductance and temperature, resulting in signal-specific SQI for each, but with different empirically determined threshold values. After this step, our approach deviates more from Böttcher’s algorithm. In particular, Böttcher aggregates each of the three SQIs to a binary value per 1-minute window by determining whether more than 1\% of that SQI is considered on-body. Subsequently, these three 1-minute SQIs are combined into a single binary SQI that becomes on-body if at least one of the signals is on-body (OR-operation).

Conversely, our approach first ensures that the three SQIs are aligned by reindexing them to the timestamps of the skin conductance SQI signal (4 Hz). Subsequently, in line with Böttcher, the three SQIs are combined via an OR-operation. This combined SQI signal is then smoothed using a 1-minute window, factoring out brief instances of wear and non-wear misdetections. This results in our final "Wrist\_SQI", illustrated in Figure~\ref{fig:off_wrist_algorithm} (a).

Figure~\ref{fig:off_wrist_algorithm} displays a visual comparison of Böttcher’s and our non-wear detection algorithms, specifically highlighting Böttcher’s lower skin temperature sensitivity. Further implementation details of both algorithms can be found on GitHub. When tested on a consumer-grade desktop (AMD Ryzen 2600x), our refined non-wear detection pipeline demonstrated an inference time of 6ms per hour of E4 wearable data, which is a substantial improvement from the 38 ms required by Böttcher’s algorithm (Table~\ref{tab:tab_non_wear}). This is especially relevant for longitudinal studies where such algorithms are applied to months of data for numerous participants. Furthermore, Supplemental S1 assesses the non-wear detection accuracy of both algorithms using a labeled subset from the mBrain dataset.

\subsection*{Challenge 6: Wearable Artifacts}
Wearable artifacts, which cause spurious signal values, primarily result from either external factors (e.g., humidity) or sensor degradation. Unlike the controlled conditions within laboratory studies, ambulatory research is subject to varying external conditions, thereby requiring methodologies which identify or mitigate impacted modalities to account for such artifacts. Among these artifacts, motion-induced artifacts are the most prevalent and have a notable impact on the wearable's physiological modalities, including photoplethysmography and skin conductance signals~\cite{schmidt_wearable-based_2019}. Moreover, improper use of wearable devices, often due to not adhering with the recommended wearing guidelines, can lead to artifacts, as observed by Stuyck et al. (2022)~\cite{stuyck_validity_2022}. Furthermore, sensor degradation issues, such as the polarization of skin conductance electrodes, is another major source of artifact generation~\cite{heikenfeld_wearable_2018, posada-quintero_innovations_2020}.

Since motion artifacts are a primary cause of signal corruption in wearable devices, numerous studies have turned to nighttime data as a means to mitigate this. Böttcher et al. (2022)~\cite{bottcher_data_2022} demonstrated, using data quality indices for each of the Empatica’s physiological signal modalities, that data collected between 8PM to 8AM exhibited substantially higher quality than daytime data. Siirtola et al. (2018)~\cite{siirtola_using_2018} conducted a wearable monitoring study, using the Empatica E4, on migraine patients with the objective of predicting the likelihood of a migraine attack within the next day. They explicitly relied on nighttime data to compute reliable features. Furthermore, Uchida et al. (2022)~\cite{uchida_use_2022} observed that the median skin temperature acquired during night via wrist-worn devices can indicate the fertility phase in women, demonstrating the reliable accuracy of nighttime data.

\begin{figure}[htb]
    \centering
    \includegraphics[clip,width=\columnwidth]{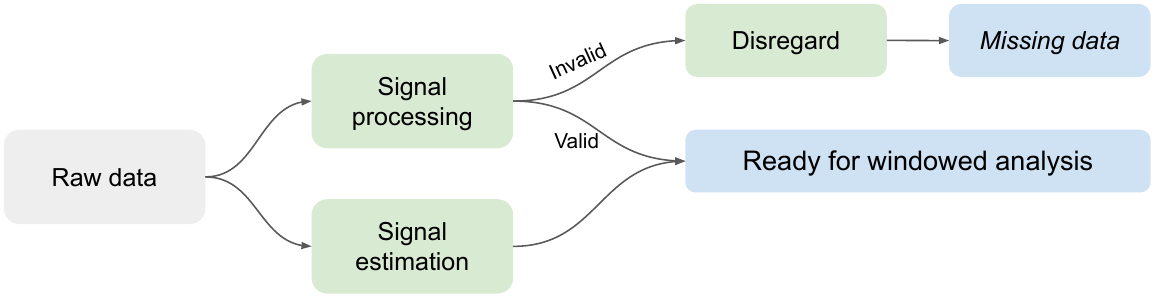}
    \caption{Flowchart for handling artifacts in raw ambulatory (daytime) wearable data.}
    \label{fig:processing_flowchart}
\end{figure}

However, the efficacy of relying solely on nighttime data may vary depending on the study’s specific objective. For instance, when the goal is to implement just-in-time interventions or to analyze physiological responses during daytime events, such as stress episodes or headaches, daytime data provides crucial information. In these cases, techniques like signal processing or signal estimation, depicted in Figure~\ref{fig:processing_flowchart}, can improve data analysis reliability.

Signal estimation leverages data-driven algorithms to enhance data quality through predicting or extracting a signal from noisy data. For instance, Reiss et al. (2019)~\cite{reiss_deep_2019} used spectral representations of the Empatica E4’s PPG signal to estimate the average instantaneous heart rate over 8-second intervals. However, a substantial constraint of signal estimation methodologies is that the majority of these are designed to replace the original signal without indicating the reliability of their estimations~\cite{moser_estimating_2017}. In contrast, signal processing refines raw signals into more usable data for further analysis. Unlike signal estimation, signal processing often includes validity scores and is generally more interpretable. Consequently, our research emphasizes the visual application and analysis of signal processing techniques.

\begin{figure}[!htb]
    \centering
    \includegraphics[clip,width=\columnwidth]{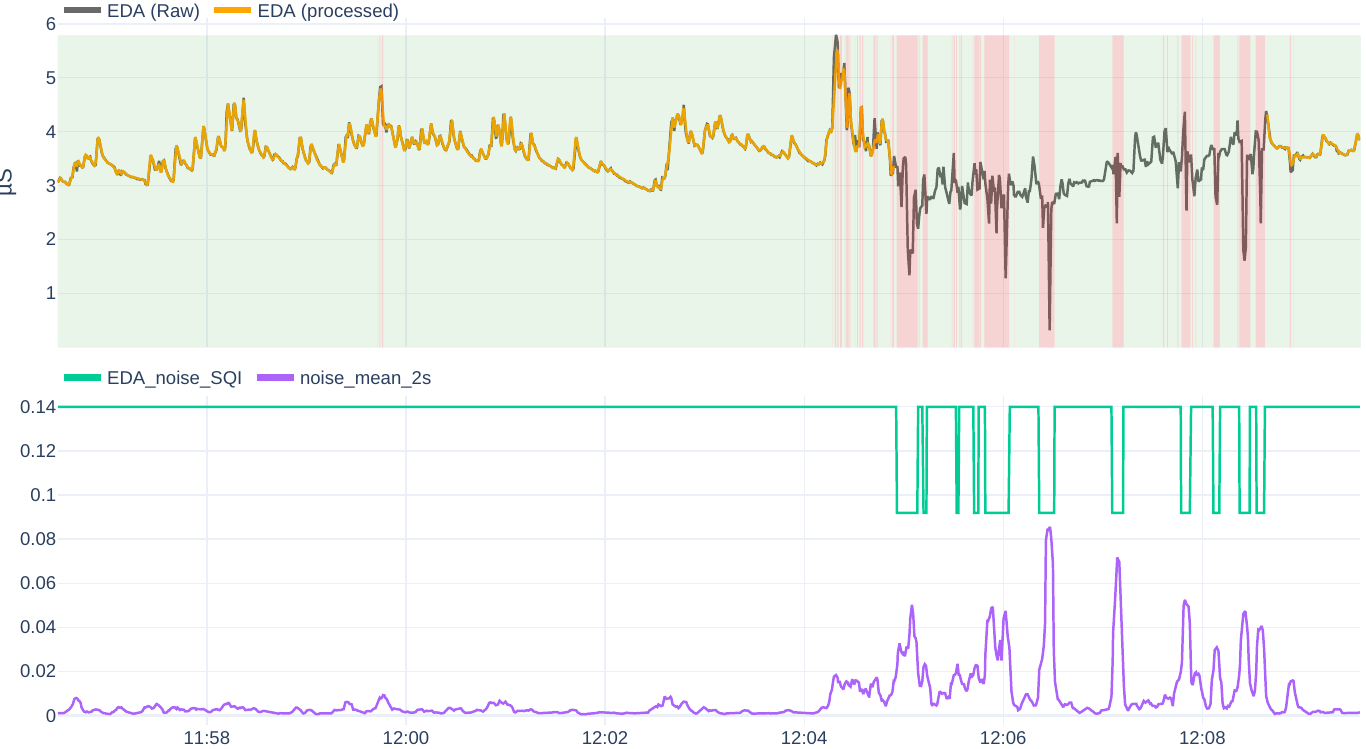}
    \caption{Flowchart for handling artifacts in raw ambulatory (daytime) wearable data. \textit{Note}:  The figure consists of two vertically stacked subplots that share the same x-axis. The upper subplot displays the raw EDA signal depicted by the gray line, with valid and invalid SQI regions distinguished by green and red backgrounds, respectively. The processed EDA data is illustrated by an orange line. Remark that there is no one-on-one relationship between the processed EDA data and the valid regions. This is because the duration and frequency of these invalid regions affect the eventual retention of the raw EDA signal. Specifically, brief and infrequent invalid segments, like those until 12h05, can be effectively imputed using interpolation, resulting in no data exclusion in the processed EDA signal. Conversely, as the frequency and/or duration of invalid segments increases, evidenced between 12h05 and 12h06, successful interpolation is compromised, resulting in disregarding these invalid regions. Moreover, processed EDA segments, but shorter than 60 seconds (e.g., valid segments between 12h06 and 12h08), are excluded given their limited analytical utility. The lower subplot elucidates the components of the skin conductance SQI. In alignment with the non-wear detection pipeline, multiple sub-SQIs are utilized. The noise amplitude of the EDA, averaged over a two-second window, is delineated by a purple line. This signal is thresholded to determine the noise sub-SQI, marked by the green line.}
    \label{fig:processing_example}
\end{figure}

\subsubsection*{Countermeasures}
In alignment with the non-wear detection from the previous paragraph, signal processing solutions often utilize SQIs to differentiate valid from invalid segments for the targeted modalities. Visual analytics are instrumental in shaping and evaluating these pipelines. As such, we introduce a generic visualization approach that we frequently employ. This visualization approach is materialized via a skin conductance processing use case, shown in Figure~\ref{fig:processing_example}.

Essentially, our approach utilizes multiple vertically stacked subplots, all sharing a common x-axis that denotes time. The uppermost subplot displays both the raw and processed signals, enabling a direct visual comparison. In this subplot, background shading accentuates the SQI outcome, simplifying the distinction between valid and invalid segments and their impact on the processed signal. Subsequent subplots provide insights into the components used in the processing pipeline, illuminating the composition of the final SQI seen in the upper subplot.

Notably, the visual analytics displayed in Figure~\ref{fig:processing_example} are realized by employing our widely-adopted open-source Python tools. The processing pipeline is constructed using tsflex, an efficient toolkit which offers functionality to wrap and serialize data processing functions for time series data, facilitating convenience and easy deployment~\cite{van_der_donckt_tsflex_2022}. The visualization is rendered with Plotly-resampler, a highly scalable time series visualization tool, which facilitates back-testing on large amounts of data~\cite{van_der_donckt_plotly-resampler_2022}. It is this interplay between efficient signal processing and scalable interactive visualization that drives thorough analysis and broad exploration on large data volumes~\cite{bernard_visual-interactive_2012}.

\subsection*{Challenge 7: Missing Wearable Data}
In ambulatory studies, encountering missing segments of wearable data is inevitable. On the one hand, missing data can simply stem from intervals during which the device is not worn, whereas on the other hand, missing wearable data periods can be introduced during signal processing (see Figure~\ref{fig:processing_flowchart}). In addition to these two sources of missing data, device particularities can also contribute to missing data. For instance, during the mBrain study, we suffered from the absence of an automatic reconnection functionality for the E4 when streaming data. Consequently, any disruption in Bluetooth connectivity, possibly caused by distancing from the paired smartphone, resulted in the E4 device shutting down. Users were then required to notice this, manually restart the device, and reconnect it to the smartphone. This limitation substantially reduced the wearable data volume compared to on device logging, even further compounded by the increased battery consumption due to Bluetooth streaming~\cite{bottcher_data_2022}.

Given the inevitability and the high prevalence of missing data in ambulatory wearable studies, the impact of missing wearable data on study results should be considered.

\begin{figure}[htp]
    \centering
    \includegraphics[clip,width=0.9\columnwidth]{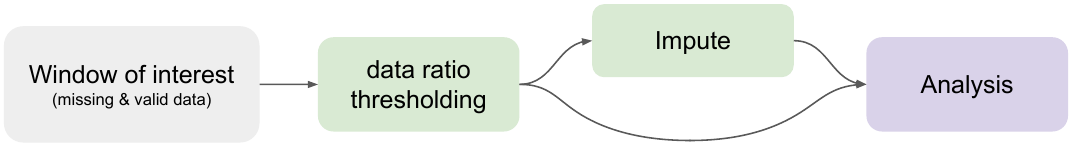}
    \caption{Flowchart illustrating the methodology for performing data analysis using incomplete data.}
    \label{fig:imputation_flowchart}
\end{figure}

\subsubsection*{Countermeasures}
\begin{figure}[htp]
    \centering
    \includegraphics[clip,width=0.65\columnwidth]{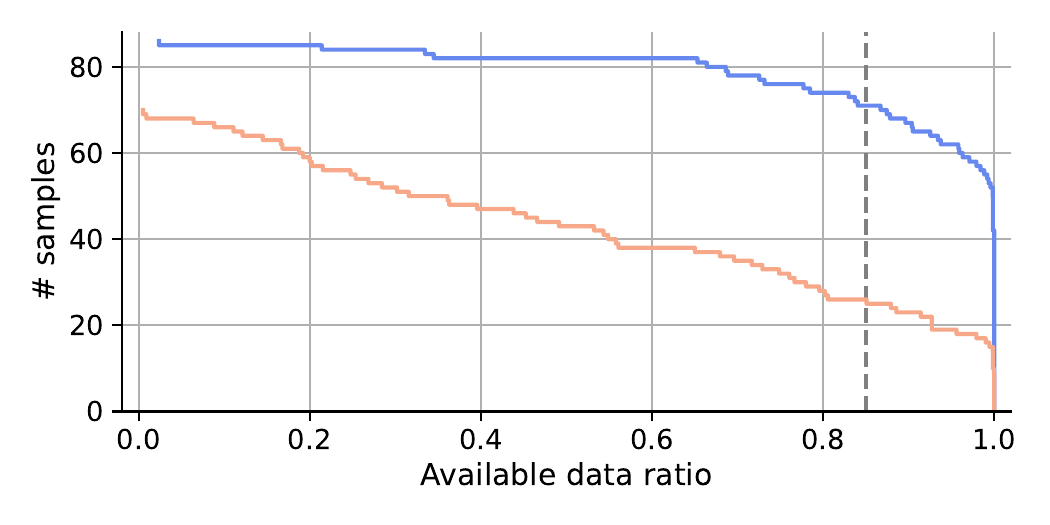}
    \caption{Complementary cumulative distribution plot of the window-of-interest data ratios for two participants. \textit{Note}: The y-axis represents the number of available window-of-interest samples, while the x-axis indicates the corresponding data ratio. Each curve in the plot represents the complementary cumulative distribution of a participant, providing a visual assessment of overall data-availability per participant. Furthermore, when utilizing a data-ratio threshold, exemplified via the dashed gray line for the data-ratio of 0.85, this visualization allows determining the remaining number of samples adhering to this threshold.}
    \label{fig:complementary_data_ratio}
\end{figure}

In the previous subsection, we outlined how signal processing and signal estimation methodologies can be utilized to address spurious data segments, leading to enhanced or excluded segments. The resulting processed data is then suitable for qualitative analysis. As depicted in Figure~\ref{fig:imputation_flowchart}, we particularly focus on “window-of-interest”-based analyses. This approach is widely employed in wearable monitoring research to partition the collected data into windows for which the analysis will take place~\cite{siirtola_using_2018, stubberud_forecasting_2023, bulling_tutorial_2014}. These windows-of-interest should be derived from the study’s research question. For instance, if the study aims to understand the precursors of headache attacks, the window-of-interest could encompass wearable data acquired the day before an episode took place~\cite{vandenbussche_patients_2024}.
Upon defining these “windows-of-interest”, one can evaluate the proportion of missing and valid data within them. Figure~\ref{fig:complementary_data_ratio} presents a complementary cumulative distribution plot of two participants, showcasing the window-of-interest data availability across different data ratios. From this plot, one can easily interpret the amount of samples that are available for a given data ratio (per participant).

A possible countermeasure to this missing data is imputation, which replaces missing values with aggregated imputation values, as outlined by Weed et al. (2022)~\cite{weed_impact_2022}.

\begin{figure}[ht]
    \centering
    \includegraphics[clip,width=\linewidth]{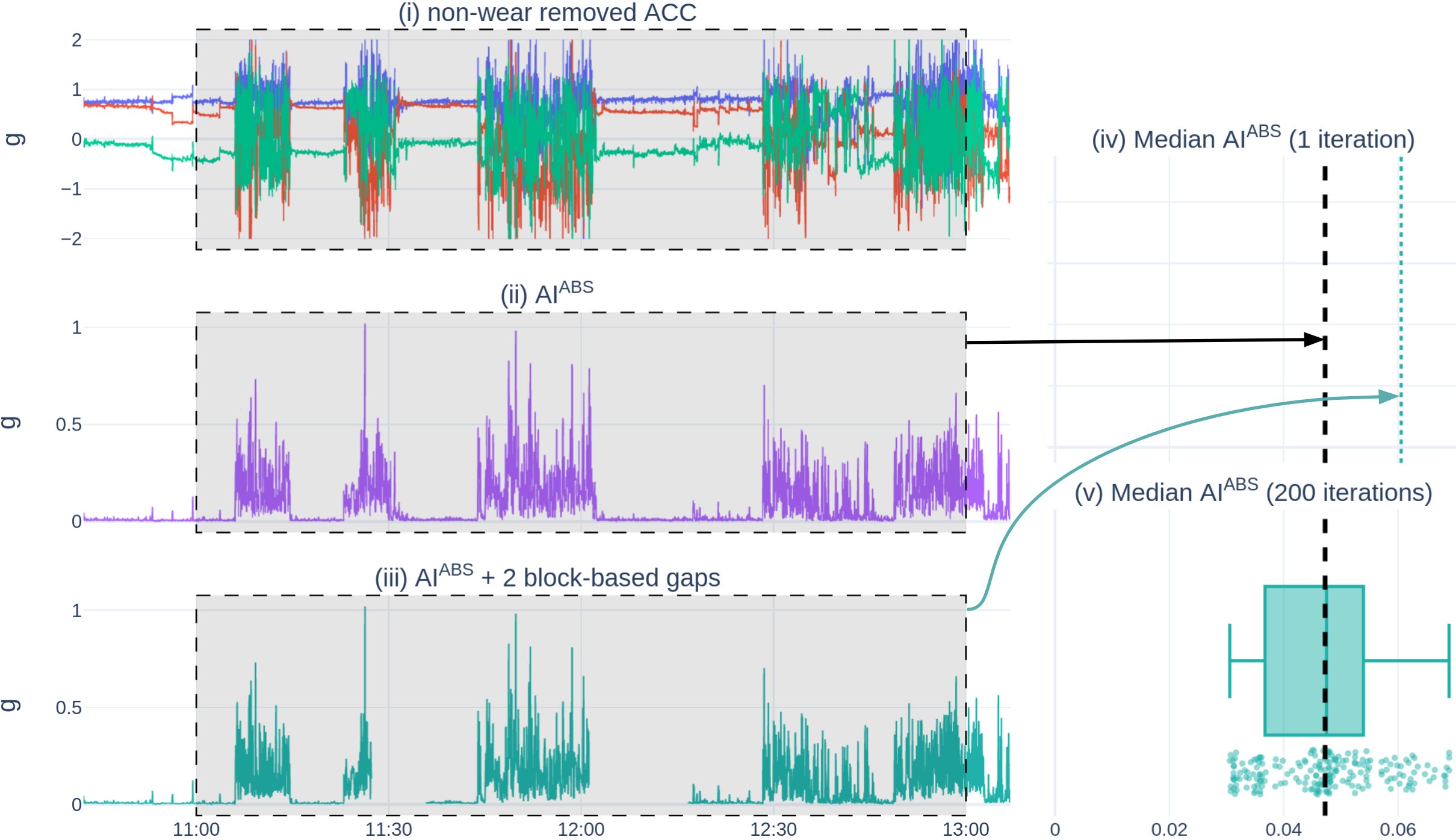}
    \caption{Overview of a single block based bootstrapping iteration using the median as desired metric. \textit{Note}: The figure comprises three vertical stacked subplots on the left that share an x-axis, with the window-of-interest highlighted by a gray shaded area. The two vertical subplots on the right side share an x-axis as well. Subplot (i) depicts an excerpt of processed wearable movement data, for which non-wear periods have been removed. Remark that no non-wear periods were detected, and no data is missing, resulting in a complete valid segment for our window of interest. Subplot (ii) visualizes the transformed ACC data of (i) into a second-per-second activity intensity index, $AI^{ABS}$, in accordance with~\cite{bai_activity_2016}. This $AI^{ABS}$ signal is then utilized to compute our desired metric values, specifically, the median value of all data within our window of interest. This reference metric value is represented by a bold dashed black line in subplot (iv) and (v). Subsequently, gap-based bootstrapping is employed utilizing the complete movement intensity data from subplot (ii) as input. In particular, one or multiple block-based gaps are generated to create a gap-induced signal, shown in subplot (iii), maintaining a specific retention data ratio, which is in this illustration 0.6. The modified signal is then used to compute the desired metric, which is depicted by the vertical green dotted line in subplot (iv). Each bootstrap iteration results in adding another data point in subplot (v), which can then be utilized to assess the spread for a given data retention ratio. Further specifics can be found on \href{https://github.com/predict-idlab/data-quality-challenges-wearables/blob/main/notebooks/mBrain/C7_missing_data.ipynb}{Github}.
}
    \label{fig:gap_imputation_methodology}
\end{figure}

To rigorously assess the impact of missing data segments on analytical results, it is generally recommended to utilize bootstrapping combined with gap simulation~\cite{berkowitz_recent_2000, efron_missing_1994}. However, such analyses are often neglected as literature tends to not consider “windows-of-interest” that contain missing data~\cite{weed_impact_2022}. We are therefore the first to introduce a detailed, step-by-step procedure for assessing the impact of missing wearable data on outcome metrics, as such allowing inclusion of “windows-of-interest” with missing data. Figure~\ref{fig:gap_imputation_methodology} illustrates this procedure using a wearable accelerometer signal.

Beginning with the window-of-interest, the first step entails selecting a processed, gap-free series and then computing the analysis metrics to obtain the \textit{reference} gap-free metric values. For a wearable movement use-case, as exemplified by Figure~\ref{fig:gap_imputation_methodology}, this step is illustrated in subplots (ii) and (iv).

The second step consists of gap bootstrapping. Specifically, one or multiple gaps are induced in order to obtain a certain data retention ratio. The gap induction method should mimic how missing data is introduced in other incomplete windows of interest. When dealing with wearable data, arbitrarily removing points is illogical. Instead, block-based gap induction methods which represent non-wear bouts are recommended~\cite{berkowitz_recent_2000}. We provide a comprehensive comparison of various gap induction methodologies applied to wearable data bootstrapping in Supplemental S2.

To facilitate statistical analysis, multiple repeats of the second step are conducted on each chosen, processed, and complete reference series. This yields a set of metric values under varying simulated gap conditions, allowing to observe the distribution and spread on the outcome metric with regard to the reference (gap-free) metric value, as shown in subplot (v) of Figure~\ref{fig:gap_imputation_methodology}.

\begin{figure}[ht]
    \centering
    \includegraphics[clip,width=\linewidth]{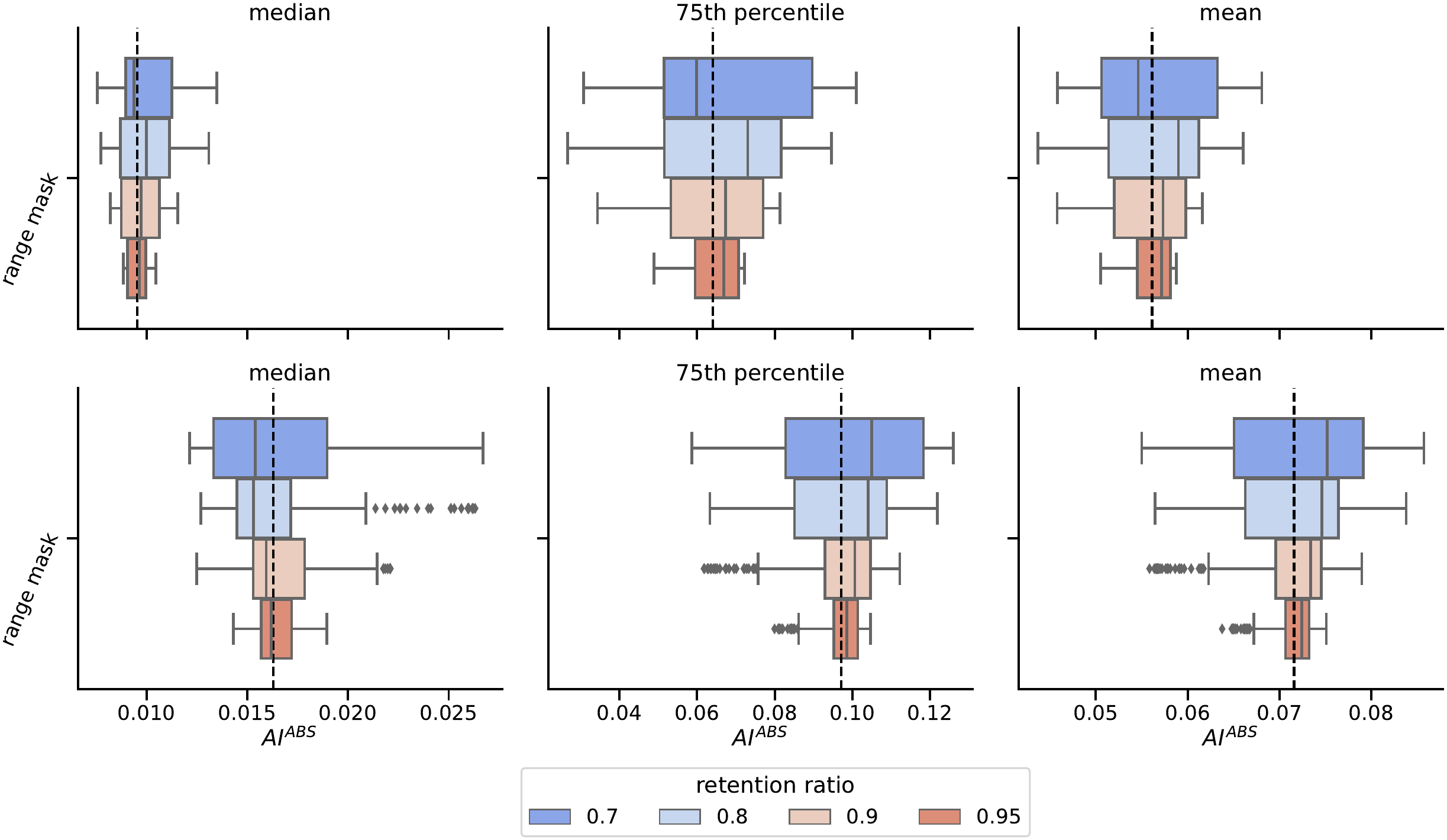}
    \caption{Spread analysis of block-based gap bootstrapping for various data ratios and metrics. \textit{Note}: Each row in the figure represents a distinct reference series, signifying a window of interest from a unique moment. Different columns correspond to varying metrics, with the vertical dashed black line illustrating the metric value of the gap-free reference series. In creating this specific visualization, the accelerometer data from the Empatica E4 was transformed into a second-by-second activity index, AIABS, as per the methodology detailed by (Bai et al., 2016) and illustrated in Figure 14. The considered metrics are the 50th percentile, 75th percentile, and mean values calculated from the $AI^{ABS}$ data of the selected time window.}
    \label{fig:bootstrap_spread}
\end{figure}

When gaps of varying data retention ratios are simulated, we can explore the impact of the data retention ratio on metric variability. Swarm plots or box plots can visualize this by depicting distributions for each combination of metric, data retention ratio, and reference series, as depicted in Figure ~\ref{fig:bootstrap_spread}. Used in conjunction with the cumulative data-ratio plot in Figure~\ref{fig:complementary_data_ratio}, this facilitates data-driven decisions regarding the data retention ratio threshold for “windows-of-interest”.

Remark that the proposed bootstrapping analysis comes with certain limitations. One significant drawback is that the gap induction procedure does not account for potential biases related to the specific time periods when participants are not wearing the watch, as outlined in Challenge 4 (\textit{bias}). Conversely, solely including complete windows of interest, which is common practice in literature, may also introduce certain biases in the downstream analysis.

Furthermore, these analyses are more representative when all “windows-of-interest” occur at fixed time spans (e.g., the wakeful period from 10AM to 8PM on the day prior to a headache event), instead of varying day-time ranges (e.g., three hours before a stress event). This is because the time-of-day influences the occurrence and nature of gaps in the data, as indicated by Weed et al. (2022)~\cite{weed_impact_2022}.

In the above paragraphs, we assessed the impact of missing wearable data during “windows-of-interest” on the outcome metric. However, another approach to overcome this missing data challenge is by utilizing imputation techniques. Recently, Weed et al. (2022)~\cite{weed_impact_2022} assessed the impact of the timing and duration of missing wearable actigraphy data on two outcome metrics for several imputation methods. Their findings showed that median time of day imputation produced the least deviation among the imputation methods. Remark that the impact of such imputation methods can also be analyzed via our proposed bootstrapping spread analysis methodology.

Furthermore, if you are working with temporal cyclical data, such as circadian dependent data, cosinor based rhythmometry may prove useful as this is a gap-robust methodology that can deal with imbalance~\cite{cornelissen_cosinor-based_2014, moskon_cosinorpy_2020}. Lastly, It is also advisable to consult literature to cross-reference the data-ratios employed in prior research, if available.

\subsection*{Summary}
To conclude the wearable data quality section, we summarize the presented three challenges and their countermeasures in Table~\ref{tab:summary_wearable}.

\begin{table}[]
\centering
\footnotesize
\begin{tabular}{@{}ll@{}}
\toprule
\textbf{Challenge} & \textbf{Countermeasures / Actions} \\ \midrule
\textbf{Wearable non-wear} &  \\
 & Perform non-wear detection as a preprocessing (data filtering) step \\ \midrule
\textbf{Wearable artifacts} &  \\
\textbf{} & \begin{tabular}[c]{@{}l@{}}steer clear off: Utilize nighttime data (overall higher data quality)\\ Signal processing: discern validity of signals (\& enhance)\\ Signal estimation: estimate target signal\\ Visual analytics of signal processing and estimation steps are crucial\\ for quality assessment\end{tabular} \\ \midrule
\textbf{Missing and Spurious data} &  \\
\textbf{} & \begin{tabular}[c]{@{}l@{}}Visualize available (processed) data retention ratios for participants\\ Computing metrics with gaps or imputation: Bootstrapping techniques \\ aid in assessing the spread of your outcome metric for a given data-ratio\end{tabular} \\ \bottomrule
\end{tabular}
\caption{Summary of wearable data quality challenges and their countermeasures.}
\label{tab:summary_wearable}
\end{table}

\section*{Limitations}
In this work, we focused on addressing seven data quality challenges in ambulatory wearable monitoring studies, stemming from participants, monitoring applications, and wearable devices.

While we touched upon participant and application-related aspects, including user burden and application experience, we did not delve extensively into psychological dimensions. For example, an important psychological aspect that we did not touch upon is intrinsic motivation, as this significantly influences study engagement~\cite{chalofsky_meaningfulness_2009}. Therefore, psychological aspects should also be considered in longitudinal studies.

Regarding wearable-related challenges, we focused on introducing innovative methodologies that target wearable data quality. This is particularly relevant in the context of daytime-based analyses, as Böttcher et al. (2022)~\cite{bottcher_data_2022} highlighted suboptimal biosignal data quality in wrist-worn wearables during daylight hours. However, our wearable-related countermeasures are not devoid of limitations. For instance, both the ETRI and mBrain datasets rely on the Empatica E4 wearable device, constraining our analytical examples to a single device. While we believe that most of our countermeasures are wearable agnostic, device-specific characteristics might affect data quality and the subsequent analyses. Future research should extend our methodologies to a diverse range of wearable devices, encompassing smartphones and chest-strap wearables.

Another limitation is that we did not delve into wearable synchronization, since only a single wearable was utilized in both studies under consideration. Within the mBrain study, the Empatica device is connected to the phone, whose timestamp is used to synchronize the Empatica, thus mitigating the smartphone and wearable synchronization challenge. However, this challenge is not neglected in literature, as the work of Wolling et al. (2021)~\cite{wolling_pulsync_2021} concentrated on providing a methodology to perform device synchronization when operating with multiple wearable devices that share a highly-correlated signal, such as heart rate.

We also refrained from discussing the measurement sensitivity of certain wearable device types. If, for instance, the objective of an ambulatory wearable study is to investigate activity patterns in participants, wrist-worn devices tend to be less accurate than chest or hip-based wearables in capturing Activity Energy Expenditure (AEE)~\cite{van_remoortel_validity_2012}. Milstein et al. (2020)~\cite{milstein_validating_2020} specifically evaluated the reliability of the Empatica E4’s skin conductance signal using the MindWare Mobile Impedance Cardiograph device to acquire palm skin conductance data as reference. Their results concluded that the E4 was not able to produce reliable EDA data, which may be attributed to lower sweat gland density on the wrist compared to the hand palm~\cite{asahina_sweating_2015}. Therefore, it is paramount during study design to first consult literature regarding the measurement sensitivity and limitations of your device at hand.

In summary, while our research offers valuable insights and methodologies for improving wearable data quality, it is imperative to consider its limitations and the need for future research to validate and extend its applicability and robustness.

\section*{Conclusion}
Recent advancements in wearable sensing technologies, particularly wrist-worn devices, offer promising solutions for longitudinal follow-up of chronic patients by shifting from intermittent, subjective self-reporting to objective, continuous monitoring. However, collecting and analyzing wearable data in conjunction with health-related records, presents unique challenges. We distinguished two categories of data-quality challenges; (i) participant- and data-entry related challenges, and (ii) wearable-related analysis challenges.

For every identified challenge, we provided insights into the causes, effects, and countermeasures. Particularly, we built upon our first-hand experience gathered during the mBrain study and utilized two public real-world datasets to illustrate both the challenges and the proposed countermeasures. This way, our work aimed to practically address the overlooked challenges in data-collection and retrospective analysis in ambulatory wearable monitoring studies.

Regarding the participant- and data-entry related challenges, a key overarching conclusion is that any component requiring user interactions should be intricately tied to the research objective and demand minimal user effort~\cite{schmidt_wearable-based_2019, schmidt_labelling_2018}. The selected wearable device should also align with this research goal in terms of measuring sensitivity and user burden. Particularly, minimizing user burden of the wearable device is paramount in longitudinal research settings~\cite{balbim_using_2021}. Furthermore, participant compliance can be monitored via compliance visualizations which leverage near real-time participant data-streams. Such visualizations can be utilized to timely re-instruct participants. In order to mitigate implicitness assumptions and minimize the likelihood of data entry errors, it is advisable to conduct monitoring studies in incremental waves, starting with a pilot study. Moreover, questionnaires can serve as a means to also factor out implicitness assumptions. Finally, incorporating tailored questionnaires which gauge for context, can aid in assessing personal bias for highly subjective event labels such as stress.

Turning to wearable-related data quality challenges, visualization plays a critical role in evaluating the quality of different signal modalities during data processing and analysis steps. The utilization of tools like tsflex and Plotly-Resampler can significantly enhance the ability to process and visualize these data modalities in an efficient and scalable fashion. Moreover, we introduce an algorithm that performs better in both inference speed and accuracy for identifying non-wear periods, which was developed using the aforementioned two toolkits. A non-wear detection pipeline is crucial to filter out non-wear bouts before conducting further processing and analysis. Finally, we propose a bootstrapping methodology for assessing the impact on the study analysis metrics of incorporating incomplete windows-of-interest.

In conclusion, we present practical solutions to prominent challenges in ambulatory monitoring research, bolstering the quality and efficacy of data collection and analysis. By openly sharing our code scripts and a subset of the mBrain study data, we facilitate reproducibility and enable direct applicability in real-world settings.

\bibliography{mbrain_methods}





\section*{Author contributions statement}
\textbf{JoVDD}: Conceptualization, Methodology, Software, Validation, Data curation, Formal Analysis, Visualization, Writing - Original Draft. \textbf{NV}: Investigation, Writing - Review and Editing. \textbf{JeVDD}: Writing - Review and Editing. SC: Writing - Review and Editing. \textbf{MS}: Investigation, Writing - Review and Editing. \textbf{MDB}: Investigation, Writing - Review and Editing. \textbf{BS}: Investigation, Writing - Review and Editing. \textbf{KP}: Investigation, Supervision, Writing - Review and Editing. \textbf{FO}: Conceptualization, Funding Acquisition, Supervision, Writing - Review and Editing. \textbf{SVH}: Conceptualization, Funding Acquisition, Supervision, Writing - Review and Editing.

\section*{Additional information}
\subsection*{Data Availability Statement}
All code and a patient sample of the mBrain study are publicly available at \url{https://github.com/predict-idlab/data-quality-challenges-wearables}  and \url{https://www.kaggle.com/datasets/jonvdrdo/mbrain21/data}, respectively.

\subsection*{Competing Interests}
The authors declare no conflicts of interest.

\subsection*{Supplementary Information}
All supplementary information is available at \href{https://github.com/predict-idlab/data-quality-challenges-wearables/blob/main/Supplemental%20information.pdf}{GitHub}.

\end{document}